\documentclass[twocolumn, 10pt]{IEEEtran}
\usepackage{hyperref} 
\usepackage{graphicx}
\usepackage{amssymb}
\usepackage[cmex10]{amsmath}
\newcommand{\Eb}[1]{{ \mathbb{E}\left[ #1 \right] }}
\newcommand{\V}[1]{{ \text{var}\left[ #1 \right] }}
\DeclareMathOperator{\I}{\mathcal{I}}

\DeclareMathOperator{\Prb}{\mathbb{P}}

\DeclareMathOperator{\Z}{\mathbb{Z}}
\DeclareMathOperator{\E}{\mathbb{E}}

\newcommand{\Tcs}{T_{\text{c}}}
\newcommand{\Tmec}{T_{\text{mec}}}

\newcommand{\pso}{p_{\text{secp}}}
\newcommand{\pco}{p_{\text{comp}}}

\newcommand{\Lc}{\lambda_{\text{c}}}
\newcommand{\Lm}{\lambda_{\text{m}}}
\newcommand{\Tc}{T_{\text{comp}}}
\newcommand{\poutul}{p_{\text{o,ul}}}
\newcommand{\poutdl}{p_{\text{o,dl}}}

\newcommand {\Define} {\stackrel {\Delta} {=}  }

\newcommand{\mya}{\mathrel{\overset{\makebox[0pt]{{\tiny(a)}}}{=}}}

\newcommand{\myb}{\mathrel{\overset{\makebox[0pt]{{\tiny(b)}}}{=}}}

\newcommand{\myc}{\mathrel{\overset{\makebox[0pt]{{\tiny(c)}}}{=}}}

\newcommand{\myd}{\mathrel{\overset{\makebox[0pt]{{\tiny(d)}}}{=}}}

\newcommand{\gleq}{{{{k_0 = \lfloor \zeta(R) \rfloor}\atop \geq}\atop <}}

\usepackage{bm}
\usepackage{mathtools}
\usepackage{epsfig,makeidx,color}
\usepackage{cite}
\usepackage{stfloats}
\usepackage{subfigure}
\usepackage{psfrag}
\usepackage[mathscr]{euscript}
\usepackage{chngcntr}
\usepackage{lineno}
\usepackage[T1]{fontenc}
\usepackage{microtype}
\usepackage{wasysym}
\usepackage{footnote}
\usepackage{acronym}  
\usepackage{booktabs} 
\usepackage[table]{xcolor}
\usepackage[table]{xcolor}
\usepackage{amsbsy}
\usepackage{amssymb}
\usepackage{psfrag}
\usepackage[mathscr]{euscript}
\usepackage{lipsum}
\usepackage{cite}
\usepackage[ruled,linesnumbered,resetcount, norelsize]{algorithm2e}
\usepackage{nicefrac}
\usepackage[english]{babel}
    \usepackage[font=small,labelsep=period]{caption}
\makeatletter
\def\citenoauxwrite#1{\begingroup
\@fileswfalse
\cite{#1}\relax
\endgroup}
\makeatother

\acrodef{CCDF}{complementary cumulative distribution function}
\acrodef{CF}{characteristic function}
\acrodef{PPP}{Poisson point process}
\acrodef{CSI}{channel state information}
\acrodef{OFDM}{orthogonal frequency division multiplexing}
\acrodef{OFDMA}{orthogonal frequency division multiple access}

\acrodef{RV}{random variable}
\acrodef{i.i.d.}{independent, identically distributed}
\acrodef{PMF}{probability mass function}
\acrodef{PDF}{probability distribution function}
\acrodef{CDF}{cumulative distribution function}
\acrodef{ch.f.}{characteristic function}
\acrodef{AWGN}{additive white Gaussian noise}
\acrodef{SNR}{signal-to-noise ratio}
\acrodef{LRT}{likelihood ratio test}
\acrodef{DRT}{distance ratio test}
\acrodef{GLRT}{generalized likelihood ratio test}
\acrodef{CRLB}{Cram\'{e}r-Rao lower bound}
\acrodef{CRB}{Cram\'{e}r-Rao bound}
\acrodef{ZZLB}{Ziv-Zakai lower bound}
\acrodef{ZZB}{Ziv-Zakai bound}
\acrodef{LOS}{line-of-sight}
\acrodef{ToF}{time-of-flight}
\acrodef{NLOS}{non-line-of-sight}
\acrodef{GDOP}{geometric dilution of precision}
\acrodef{GPS}{Global Positioning System}
\acrodef{FIM}{Fisher information matrix}
\acrodef{PEB}{position error bound}
\acrodef{SPEB}{squared position error bound}
\acrodef{TOA}{time-of-arrival}
\acrodef{TOF}{time-of-flight}
\acrodef{WSN}{wireless sensor network}
\acrodef{MAC}{medium access control}
\acrodef{RSS}{received signal strength}
\acrodef{WAF}{wall attenuation factor}
\acrodef{TDOA}{time difference-of-arrival}
\acrodef{RF}{radiofrequency}
\acrodef{RTT}{round-trip time}
\acrodef{AOA}{angle-of-arrival}
\acrodef{MF}{matched filter}
\acrodef{ED}{energy detector}
\acrodef{ML}{maximum likelihood}
\acrodef{MSE}{mean-square error}
\acrodef{RMSE}{root-mean-square error}
\acrodef{LEO}{localization error outage}
\acrodef{ppm}{part-per-million}
\acrodef{ACK}{acknowledge}
\acrodef{UWB}{Ultrawide bandwidth}
\acrodef{TNR}{threshold-to-noise ratio}
\acrodef{LS}{least squares}
\acrodef{IR-UWB}{impulse radio UWB}
\acrodef{FCC}{Federal Communications Commission}
\acrodef{TH}{time-hopping}
\acrodef{PPM}{pulse position modulation}
\acrodef{MUI}{multi-user interference}
\acrodef{PDP}{power delay profile}
\acrodef{BPZF}{band-pass zonal filter}
\acrodef{SIR}{signal-to-interference ratio}
\acrodef{RFID}{radio frequency identification}
\acrodef{WPAN}{wireless personal area network}
\acrodef{WWB}{Weiss-Weinstein bound}
\acrodef{DP}{direct path}
\acrodef{MF}{matched filter}
\acrodef{MMSE}{minimum-mean-square-error}
\acrodef{SBS}{serial backward search}
\acrodef{SBSMC}{serial backward search for multiple clusters}
\acrodef{NBI}{narrowband interference}
\acrodef{WBI}{wideband interference}
\acrodef{INR}{interference-to-noise ratio}
\acrodef{CR}{channel response}
\acrodef{CIR}{channel impulse response}
\acrodef{CR}{channel  response}
\acrodef{RADAR}{radar}
\acrodef{MUR}{Multistatic radar}
\acrodef{JBSF}{jump back and search forward}
\acrodef{HDSA}{high-definition situation-aware}
\acrodef{RRC}{root raised cosine}
\acrodef{ST}{simple thresholding}
\acrodef{BTB}{Bellini-Tartara bound}
\acrodef{P-Max}{$P$-Max}  
\acrodef{MIMO}{multiple-input multiple-output}
\acrodef{MAP}{maximum a posteriori}
\acrodef{FG}{factor graph}
\acrodef{OP}{outage probability}
\acrodef{WED}{wall extra delay}
\acrodef{RMS}{root mean square}
\acrodef{SPAWN}{sum-product algorithm over a wireless network}
\acrodef{MDD}{minimum distance distribution}
\acrodef{MAP}{maximum a posteriori probability}
\acrodef{PAR}{probabilistic association rule}

\usepackage{color}
\usepackage{dsfont}
\usepackage{bbm}

\newcommand{\Ws}[2]{{W_{}^{}}} 

\newcommand{\TSIR}[2]{{\tau_{}^{}}}

\DeclareMathAlphabet{\mathsf}{OML}{cmbr}{m}{it}

\newtheorem{theorem}{Theorem}

\newtheorem{corollary}{Corollary}
\newtheorem{proposition}{Proposition}

\newtheorem{remark}{Remark}

\newcommand{\HGF}[3]{{}_{#1}F_{#2}\!\left(#3\right)}

\newcommand{\bd}{\begin{description}}
\newcommand{\ed}{\end{description}}
\newcommand{\be}{\begin{enumerate}}
\newcommand{\ee}{\end{enumerate}}
\newcommand{\bi}{\begin{itemize}}
\newcommand{\ei}{\end{itemize}}
\newcommand{\bl}{\begin{list}}
\newcommand{\el}{\end{list}}
\newcommand{\bt}{\begin{tabbing}}
\newcommand{\et}{\end{tabbing}}

\acrodef{BS}{base station}
\acrodef{AP}{access point}
\acrodef{TDD}{time-division duplexing}

\begin{document}

\newcommand{\paperTitle}{Edge Computing-Enabled \\Cell-Free Massive MIMO Systems}
%
%
 
 

\title{\paperTitle}

\author{
	\vspace{0.2cm}
        Sudarshan~Mukherjee, \textit{Member, IEEE} and 
        Jemin~Lee, \textit{Member, IEEE}
%
%
%
    \thanks{
        S.  Mukherjee and J. Lee are with    
       the Department of Information \& Communication Engineering (ICE), DGIST, Republic of Korea      
       (e-mail:\texttt{smukho17@dgist.ac.kr}, \texttt{jmnlee@dgist.ac.kr}). 
       A part of the material presented in this paper was presented in 2018 IEEE Globecom workshop \cite{GC18}.
}
    \thanks{
       The corresponding author is J. Lee. 
       }
}

\maketitle 

%

%

%
\setcounter{page}{1}
\acresetall
\begin{abstract}
Mobile edge computing (MEC) has been introduced
to provide additional computing capabilities at network edges
in order to improve performance of latency critical applications.
In this paper, we consider the cell-free (CF) massive MIMO
framework with implementing MEC functionalities. We consider
multiple types of users with different average time requirements
for computing/processing the tasks, and consider access points
(APs) with MEC servers and a central server (CS) with the cloud
computing capability. After deriving successful communication
and computing probabilities using stochastic geometry and queueing theory, we present the successful edge computing probability
(SECP) for a target computation latency. Through numerical
results, we also analyze the impact of the AP coverage and the
offloading probability to the CS on the SECP. It is observed that
the optimal probability of offloading to the CS in terms of the
SECP decreases with the AP coverage. Finally, we numerically
characterize the minimum required energy consumption for
guaranteeing a desired level of SECP. It is observed that for
any desired level of SECP, it is more energy efficient to have
larger number of APs as compared to having more number of
antennas at each AP with smaller AP density.
\end{abstract}


\acresetall


\section{Introduction}
Over the past few years, there has been a rapid increase in computationally intensive applications, e.g., virtual reality, autonomous driving, traffic control etc., which has given rise to the demand for additional computation resources \cite{IMT2020}. Cloud computing in radio access networks (C-RAN) was introduced as a means to cater to the need for additional computing resources \cite{Lataief2}. However, C-RAN has been known to have a centralized architecture, where additional computing resources are placed at the core network server. This increases the transmission latency for network edge users, which essentially makes the cloud computing for those users a less viable option. In recent years, to solve this problem, a new paradigm has been introduced, where the cloud computing capabilities are delegated among the network edge servers \cite{Chiang}. This new paradigm is known as mobile edge computing (MEC) \cite{Pompili}.

In MEC, the servers are placed in the proximity to the users, expecting to satisfy the critical latency requirements of computationally intensive applications in the fifth generation (5G) wireless systems. {Recently, several works have been published in MEC-enabled networks, addressing the issue of latency minimization, task offloading, resource management, and energy consumption such as \cite{Lataief6,Quek, Kilper, Youkim, Huning, Magur, Xiaoyang}.} 

{From the edge computing system design point of view, {it is observed that the works in \cite{Lataief6,Quek, Kilper, Youkim, Huning, Magur, Xiaoyang} limited their system model to either a single/known number of edge computing servers and a predetermined number of offloaded tasks/users.
They also further simplified the communication model 
by considering the average channel gain \cite{Lataief6, Huning, Magur, Xiaoyang} or constant channel gain \cite{Quek}, Youkim,
or by ignoring the co-channel interference from other base stations \cite{Lataief6,Quek, Kilper, Youkim}.
Such simplifying assumptions, however, limit the applicability of their results to a realistic large scale edge computing enabled networks.}}

{From the perspective of energy consumption, it is observed that most of these works adopt a simplified version of communication energy consumption, {by modeling} the transmission energy consumption only \cite{Lataief6,Quek, Kilper, Youkim, Huning, Magur, Xiaoyang}. {However, the energy consumption of various signal processing at the transmitter and receiver is not negligible, compared to the transmission energy consumption.}
Hence, a more realistic and detailed energy consumption model needs to be considered in the MEC scenario.}

{From the perspective of computation latency model, the computation latency model used in most of these works follow a deterministic model and therefore fail to capture the effect of randomness in computation latency (i.e., waiting and service time in a queue). }
In the recent years, some of the works have attempted to address this issue by considering queueing model for task execution at the servers. {For instance,} \cite{Niu} considers the minimization of the average computation latency, while \cite{Huang} analyzes the trade-off between the average computation latency and the network connectivity. In \cite{Lee}, the authors analyze the probability of successful computation for MEC-enabled heterogeneous networks. However, none of these works consider the impact of the successful computation probability on the energy consumption, which is also one of the key performance metrics in 5G systems. 


{{In MEC, a user offloads its task to a MEC server in the uplink, and the processed data needs to be back to the user in the downlink. Hence, the success of edge computing operation also depends on the communication performance.  
With the advent of the fifth generation (5G) wireless standards, new high performing technologies have been introduced. One such key technology is the massive MIMO system \cite{Andrews}, which is being increasingly adopted in different networking and computing frameworks (e.g., C-RAN etc.) due to its large gain in energy and spectral efficiency \cite{Dichen, Bursa}.} Inspired by these applications, in this paper, we adopt the massive MIMO framework for implementing edge computing.  
}

To fully utilize the benefits of massive MIMO in an edge computing scenario, we consider the cell-free (CF) massive MIMO framework, which is the network-centric version of massive MIMO systems \cite{Ngo1}. 
The CF massive MIMO supports an antenna system, where antennas are distributed over multiple access points (APs), located throughout the network, and these APs maintain their coordination with a central server (CS) via reliable backhaul links.{\footnote[1]{ {The CF massive MIMO system is often termed as the ``network MIMO'' system due to its distributed antenna structure \cite{Chen}, which is similar to the C-RAN. However, the CF massive MIMO system possesses a central server (CS), which connects all the APs in the system, via a reliable backhaul network and can perform joint processing of data in the uplink and joint transmission in the downlink, which is not possible in the conventional C-RAN.}}} 
For implementing edge computing, we consider the system, which has APs equipped with independent MEC servers and a CS with cloud computing capability. 
Each AP serves all the users within its coverage, i.e., a circle with fixed radius, and each user is allowed to avail the additional computing resources either at the CS or at one of the connected/serving APs (i.e., MEC servers) with some probability. 
Based on this framework, we analyze the successful edge computing probability (SECP) by considering both the communication and computation performances. Following this, we explore the relation between the SECP and the total network energy consumption to give an insight on the energy efficient design of the edge computing-enabled CF massive MIMO system. The main contributions of this paper can be summarized as follows:

\bi
\item We firstly analyze the uplink and downlink transmission performances of the CF massive MIMO system using stochastic geometry, with the maximum ratio combining (MRC) receiver in the uplink and maximum ratio transmission (MRT) beamformer in the downlink, and present an expression for the successful communication probability (SCMP);{\footnote[2]{{Although for brevity, we have limited our analysis to the MRC receiver in the uplink and MRT beamformer in the downlink, the performance with other forms of receivers and beamformers (e.g., MMSE etc.) can also be studied using the proposed communication strategy.}}} 

\item For the edge computing scenario, we consider multiple types of tasks, which take different average time for processing at CS/MEC servers. Using queueing theory, we then derive the successful computation probability (SCP) for a target computation latency, and analyze the impacts of various system parameters (e.g., AP coverage radius, the offloading probability to the CS, and AP density) on the SCP;

\item Following the analysis of communication and computation performances, we finally evaluate the successful edge computing probability (SECP) and explore how the coverage radius and the offloading probability to the CS affect the SECP. 
We also analyze the performance of the distributed antenna systems in terms of the SECP for different antenna densities, compared to that of co-located antenna system; and 
\item Finally, we characterize the minimum average total energy consumption of the proposed edge computing-enabled system, for guaranteeing a minimum desired SECP. The effect of AP density on the relation between the minimum desired SECP and total energy consumption is also explored.
\ei

\begin{figure}[t!]
    \begin{center}   
    { 
	 \includegraphics[width=1.00\columnwidth]{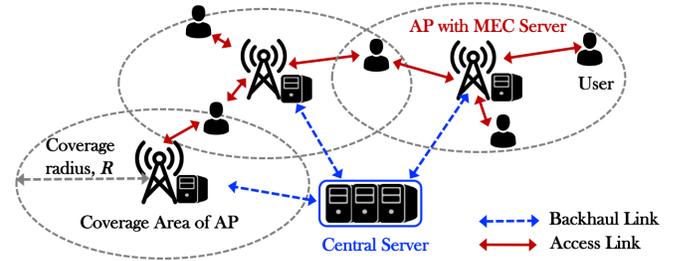}
    }
    \end{center}
    \caption{
    		An example of edge computing-enabled CF massive MIMO system. 
		 }
   \label{fig:NetworkModel}
\end{figure}

The remainder of this paper is organized as follows: Section~\ref{sec:models} introduces the proposed communication and computation model and performance metrics. In Section~\ref{sec:commperformance} and Section~\ref{sec:scp} we respectively analyze the communication and computation performances and derive an expression for the SECP. In Section~\ref{sec:results1}, we analyze the impact of various system parameters on the computation and communication performances. Finally, in Section~\ref{sec:results2}, we discuss the energy consumption parameters for the proposed edge computing scenario and analyze impact of the SECP on the total minimum required energy consumption for a minimum desired SECP. Conclusions are presented in Section~\ref{sec:conclusion}.


\section{Proposed Network Model}\label{sec:models}
In this work, we consider an edge computing enabled cell-free (CF) massive MIMO communication system, where single/multi-antenna access points (APs) are randomly distributed over the network. All these APs are interconnected via a central server (CS) through reliable backhaul links (see Fig.~\ref{fig:NetworkModel}). Note that similar to the conventional massive MIMO systems, the total number of AP antennas in a CF massive MIMO system is also significantly large (of the order of several hundreds) \cite{Ngo1}. On the other hand, the number of antennas allocated per AP, $M$, is relatively small (similar to that in the conventional small scale MIMO systems). We model the location of APs as a homogeneous Poisson point process (PPP) $\bm \Phi_b$ with density $\lambda_b$ \cite{Chen}. Similarly, the users are also assumed to be randomly located throughout the network and we also model the location of these users as a homogeneous PPP $\bm \Phi_d$ with density $\lambda_d$.

\subsection{Proposed Communication Model}
In conventional CF massive MIMO systems, a huge power gain, similar to that in conventional massive MIMO systems, is obtained by allowing each user to connect to all available APs in the system. In our proposed network model, we however consider a more practical user-centric approach, where each user can connect to only a few of the available APs. We assume that each AP has a coverage, and can reliably support all users inside this coverage circle with radius $R$ (see Fig.~\ref{fig:NetworkModel}). Hence, any user in our proposed system would be able to connect to all the APs, which are located within a distance $R$ from the user.\footnote[3]{This type of limited connectivity for users, while is more energy efficient, reduces the achievable power gain for individual users compared to that in the conventional CF massive MIMO scenario. However, in \cite{Buzzi}, it has been shown that the loss in sum-rate performance in user-centric communication approach compared to the conventional CF massive MIMO is negligible.} We assume the proposed CF massive MIMO system is operating in time division duplexed (TDD) mode and \emph{the channel between any transmitting and receiving antenna pairs are independent and Rayleigh faded}.  For the above proposed scenario, we analyze the signal-to-interference power ratio (SIR) of uplink and downlink transmissions in an interference-limited channel.{\footnote[4]{{Note that inclusion of noise in the system model would further complicate the uplink outage analysis expression, while the trends observed for the overall communication performance would be similar \cite{Ganti, Dhillon, Novlan, Nigam}. Therefore, we have limited our analysis to the interference-limited environment only.}}} 
%
%
\subsubsection{Uplink SIR Model}

Since the user distribution is assumed to be a homogeneous PPP, without loss of generality, we consider the SIR at a \emph{typical user}. We denote this user as the $0$-th user. We assume that each AP deploys a maximum ratio combining (MRC) receiver. Therefore, for the $k^{\text{th}}$ AP connected to the $0$-th user, the overall received SIR at this AP is given by 
\begin{align}
\label{eq:sirul}
\text{SIR}_{k0,\text{ul}} & = \frac{g_{k0} \ell(x_{k0})}{\sum\limits_{q \in \bm \Phi_d\backslash{\{0\}}} \widetilde{g}_{kq} \ell(x_{kq})} \, ,
\end{align}
\noindent where $g_{k0}\sim \Gamma(M,1)$ is the total received channel gain at the $k^{\text{th}}$ AP from the $0$-th user and $\widetilde{g}_{kq} \sim \text{Exp}(1)$ models the independent channel gains for the $q^{\text{th}}$ interfering user. Here, $x_{kq}$ denotes the link distance between the $q^{\text{th}}$ user and the $k^{\text{th}}$ AP and $\ell(x)$ denotes the pathloss function for a given link distance $x$.

\subsubsection{Downlink SIR Model}

{In the downlink, each AP beamforms the processed data to its associated users simultaneously using the maximum ratio transmission (MRT). Since each user receives transmission from multiple APs simultaneously, the overall received signal at the user would be sum of the beamformed data from all associated APs. This is similar to the joint transmission strategy considered in \cite{Tanbourgi}, and therefore the overall received SIR at the $0$-th user is therefore given by}
%
%
\begin{align}
\label{eq:sirdl}
\text{SIR}_{0,\text{dl}} & = \frac{ \sum\limits_{k \in \bm C(0, R) } g_{k0} \ell(x_{k0}) }{I_{\text{dl}} } \, ,
\end{align}
\noindent where $g_{k0} \sim \Gamma(M,1)$ is the channel gain received from the $k^{\text{th}}$ AP to the $0$-th user and $\bm C(0,R)$ denotes the set of APs that are connected to the $0$-th user (i.e. within a distance $R$ from the user). Here, $I_{\text{dl}}$ denotes the total interference power from all APs, which is given by 
\begin{align}
\label{eq:interfdl}
I_{\text{dl}} = \sum\limits_{k \in \bm \Phi_b} \sum\limits_{q \in \Phi_{d,k} \backslash {\{0\}} } \widetilde{g}_{kq} \ell(x_{k0}) \,.
\end{align}
\noindent Here $\Phi_{d,k}$ denotes the set of users that are connected to the $k^{\text{th}}$ AP and $\widetilde{g}_{kq}$ is $\text{Exp}(1)$ distributed and independent power gain from the beamformed signal for the $q^{\text{th}}$ interfering user.
\subsubsection{Pathloss Model}
{The singular pathloss model is defined as $\ell(r) = r^{-\alpha}, \, \, \forall \, r \geq 0$, for the link distance $r$ and the pathloss exponent of the channel $\alpha ( > 2)$, and it has been widely used in the network performance analysis such as \cite{Jlee,Ganti}. However, from \eqref{eq:sirul} and \eqref{eq:interfdl}, it is clear that the overall mean interference power would diverge with the singular pathloss model. To avoid this divergence, we consider a more accurate non-singular pathloss function,\footnote[5]{{The non-singular pathloss function has been considered as a more accurate model as it gives finite mean interference power, especially when interferers can exist near the receiver with the distance $r \geq 0$, the same as our model \cite{Bai}.}} which is given by \cite{Haenggi, Ganti2}}
{\begin{align}
\label{eq:pathloss}
\ell(r) & = \max(r, d_0)^{-\alpha} \, \, \forall \, r \geq 0 \, ,
\end{align}
where $d_0 >0$ models the reference distance between any transceiver pair.}
\subsection{Proposed Task Offloading Model}
{We consider APs which are equipped with independent MEC servers and the CS also has cloud computing capability. Thus, the task offloaded by any user can either be processed at the CS or at the MEC servers connected to the user. We also assume that the probability that the task is processed at the CS is defined as $\vartheta$ and the probability that a user chooses the connected MEC servers for processing of the offloaded task is $(1-\vartheta)$.\footnote[6]{{The offloading decision in this paper is made in a decentralized way
without the information on the current loads at CS or all other MEC servers, which is practically difficult to know at each AP. 
The offloading probability $\vartheta$ can be determined to minimize the average computation latency, which is provided in Section IV of this paper. 
Note that with the load information of all MECs, connected to the CS, 
a task scheduling mechanism at the CS can also be developed. 
However, this is not handled in this paper as it is out of the scope.}} We denote the total computation latency for MEC server processing as $\Tmec$ and for processing at the CS as $\Tcs$. Therefore, the overall computation latency for the proposed task offloading model is given by}
{\begin{align}
\label{eq:complat}	
\Tc & =	\left \{ \begin{array}{ll}
\Tcs, & \text{with probability } \vartheta\\
\Tmec, & \text{with probability } 1 - \vartheta
\end{array}\right. \, .
\end{align}}
\subsubsection{{Computing Model at the MEC Servers}}
{Note that in our proposed edge computing model, each user is connected to multiple MEC servers. Therefore, if a user chooses to process the offloaded task at a MEC server, the task processing can be done at one of the connected MEC servers. We assume that the MEC server which processes the offloaded task is selected based on the minimum instantaneous computation load. We term this MEC computation model as the \emph{minimum load computation model} (MLCM). In this case, the total latency for processing of the task, offloaded by the user, at the MEC server is given by}
{\begin{align}
\label{eq:latmecdefnew}
\Tmec & \Define T_{\text{m}, \widehat{k}}, \,\, \text{where } \widehat{k} \Define \arg\min\limits_{k \in \{1, 2, \ldots, n\}} N_{\text{m},k} \, , 
\end{align}}
{\noindent where $T_{\text{m},k}$ denotes the computation delay and $N_{\text{m},k}$ denotes the instantaneous queue length/load at the $k^{\text{th}}$ MEC server and $n$ is the number of MEC servers connected to the user.}

\subsubsection{Task Execution Time at the Servers} 
We also consider multiple types of users, which are classified by the average time for computing/processing the task at the servers \cite{Collings, Kelly}. Specifically, for the $i$-th type of task, we assume the average task processing time is $1/\mu_{\text{c},i}$ second at the CS and $1/\mu_{\text{m},i}$ second at the MEC servers. The computing time is known as the service time in the queueing model. {As other works on edge and cloud computing in \cite{Lubin, Tian, Niu, Huang, Xiahuang, Aghvami, Feng, Correia}}, we assume the computing time to be exponentially distributed for the $i$-th type task with a service rate $\mu_{h,i}$, $\forall h = \{\text{c}, \text{m}\}$. {\footnote[7]{{It has also been shown that for data centers/cloud servers and various web-based applications (e.g., HTTP and J2EE), the exponential distribution can be a good approximation for service time model \cite{Lubin,Tian}}}} Clearly, when there are $\mathcal{I}$ types of users in the system, the overall distribution of the service time $\tau_{h}$, $\forall {h} \in \{ \text{c}, \text{m} \}$, at the CS and a MEC server are given by

%
\begin{align}
\label{eq:servtpdf}
f_{\tau_{{h}}}(t) & = \sum_{i=1}^{\mathcal{I}} \frac{p_i}{\mu_{{h},i}} e^{-\mu_{{h},i} t} \, , \hspace{0.2 cm}\,\, t \geq 0, 
\end{align}
where $p_i$ is the probability of occurrence of the $i^{\text{th}}$ type task and $\sum_{i=1}^{\I}p_i = 1$.
\subsection{Performance Metrics}

In order to analyze the computation and communication performances, we use the following performance metrics: (a) successful communication probability (SCMP), (b) successful computation probability (SCP), and (c) successful edge computing probability (SECP). In the following, we first define SCMP as a function of coverage radius $R$ and the SCP as a function of $R$ and $\vartheta$ and target computation latency $t$. Finally, using these metrics, we define SECP.
\subsubsection{Successful Communication Probability}
The overall communication is assumed to be successful if both uplink and downlink transmissions are completed successfully. Hence, we define the SCMP as
\begin{align}
\label{eq:scmpdef}
p_{\text{scmp}}(R) & \Define \{1 - \poutul(R)\} \{1 - \poutdl(R)\} \, ,
\end{align}
\noindent where $\poutul(R)$ and $\poutdl(R)$ are the outage probabilities of uplink and downlink transmission respectively, for a given coverage radius $R$.

\subsubsection{Successful Computation Probability}
It is clear that the offloaded task from any user is executed either at the CS or at the MEC servers. Using the definition of total computation latency in \eqref{eq:complat}, we define the successful computation probability (SCP) for a target latency as
\begin{align}
\label{eq:scpdef}
p_{\text{comp}}(R, \vartheta, t) & = \Prb[\Tc \leq t] \,.
\end{align}
\subsubsection{Successful Edge Computing Probability}
The SECP is defined when both the computation and communication of the offloaded task are successful. {Note that the MEC servers/APs involved in the uplink communication and those considered in the computation process are the same. Therefore, the overall SECP for the proposed edge computing scenario is defined as below}
{\begin{align}
\label{eq:secpdef}
\nonumber \pso(R, \vartheta, t) & = \sum\limits_{n=1}^{\infty} \frac{(\lambda_b \pi R^2)^n}{n!} e^{-\lambda_b \pi R^2} \Prb \left[ \Tc \leq t | N = n\right]\\
& \hspace{1 cm} \times \{1 - \poutul^{(n)}(R)\} \{1 - \poutdl(R)\} \, ,
\end{align}}
{where $\poutul^{(n)}(R) \Define \Prb \left[ \max\limits_{k \in \{1,2,\ldots, n\}} \text{SIR}_{k0,\text{ul}} < \widehat{\gamma}| N = n\right] = (1 - p_0(R))^n$ (see \eqref{eq:outageul2}) and $N$ is the number of APs/MEC servers connected to a user in uplink transmission. Here, $N$ is Poisson distributed with mean $\lambda_b \pi R^2$. Therefore, $\Prb \left[ \Tc \leq t | N = n\right]$ is given by }
{\begin{align}
\label{eq:nscp}
\nonumber \Prb \left[ \Tc \leq t | N = n\right] & = \vartheta \Prb[\Tcs \leq t]\\
& \hspace{0.5 cm} + (1 - \vartheta)\Prb[\Tmec \leq t | N = n] \, .
\end{align}}
{where $\Prb[\Tmec \leq t | N = n]$ is defined in \eqref{eq:mec1}.}

\section{Communication Performance Analysis}\label{sec:commperformance}
The overall communication performance depends on both the uplink and downlink transmissions. In this section, we first derive outage probabilities for uplink and downlink transmissions, and then analyze the overall successful communication probability (SCMP) as a function of these outage probabilities.
%
\subsection{Outage Probability for Uplink Transmission}
Note that in our proposed edge computing system, all APs are interconnected through the CS via a reliable backhaul network. Hence, APs can share the offloaded data among them after a user sends it in the uplink. Therefore, it is sufficient if at least one of the APs, connected to the user, can successfully receive the offloaded data. In other words, the uplink transmission of a user can be in outage if all the APs connected to it fail to receive the data successfully. In terms of the received SIR at the connected APs, the uplink outage probability for the $0$-th user (i.e. the reference user) is then given by
\begin{align}
\label{eq:outul}
\poutul & \Define \Prb[\text{SIR}_{k0, \text{ul}} < \widehat{\gamma}, \, \forall k \in \bm \Phi_b \cap \bm C(0,R)] \, ,
\end{align}
\noindent where $\widehat{\gamma}$ denotes the threshold SIR for successful uplink communication. Using the above definition in \eqref{eq:outul}, in the following theorem, we derive the uplink outage probability for the $0$-th user as a function of the coverage radius, $R$.
\begin{theorem}\label{trm:outageul}
The uplink communication outage probability for the $0$-th user is given by
\begin{align}
\label{eq:pout}
\poutul(R) & = \exp\left \{ -2 \pi \lambda_b \int_{0}^{R}r\sum_{m=0}^{M-1}\frac{(-1)^m}{m! \, l^m(r)} \mathcal{L}_{I}^{(m)}\Big(\frac{\widehat{\gamma}}{l(r)}\Big) \, dr \right \} \, 
\end{align}
where $\mathcal{L}_{I}(s)$ is the Laplace transform of the total interference power, $I = \sum\limits_{q \in \bm \Phi_d\backslash{\{0\}}} \widetilde{g}_{kq} \ell(x_{kq})$, and $\mathcal{L}_{I}^{(m)}(s)$, the $m^{\text{th}}$ order derivative of $\mathcal{L}_{I}(s)$, is given by
%
%
\begin{align}
\label{eq:ulltderiv}
\mathcal{L}_{I}^{(m)}(s) & = \sum\limits_{i=1}^{m}{m\choose i} F^{(i)}_1(s) \, F^{(m-i)}_2(s) \,.
\end{align}
\noindent Here, $F^{(m)}_i(s)$ ($i = 1, 2$) is the $m^{\text{th}}$ order derivative of $F^{}_i(s)$, where $F_1(s) \Define \exp\left\{- \pi \lambda_d d_0^2 \Big(1 - \frac{1}{1+s d_0^{-\alpha}}\Big) \right\}$ and $F_2(s) \Define \exp \left \{ -\frac{2\pi \lambda_d}{\alpha} \frac{s d_0^{2 - \alpha} }{1 - \frac{2}{\alpha}} \HGF{2}{1}{1,1-\frac{2}{\alpha};2-\frac{2}{\alpha};-s d_0^{-\alpha}} \right\}$. Therefore, we have
%
%
\begin{align}
\label{eq:partderiv1}
\nonumber F^{(m)}_1(s) & = \pi \lambda_d \sum\limits_{i=0}^{m-1}{m-1 \choose i} (m-i)! (-1)^{m-i} F^{(i)}_1(s)\\
& \hspace{2 cm} \times \,  \frac{(d_0^{-\alpha})^{m-i}}{(1 + s d_0^{-\alpha})^{m-i+1}} \, \\
\label{eq:partderiv2}
\nonumber F^{(m)}_2(s) & = \frac{2 \pi \lambda_d}{\alpha} \sum\limits_{i=1}^{m} {m-1 \choose m-i}(-1)^{i} i! F^{(m-i)}_2(s)\\
& \times \, \frac{d_0^{2-i\alpha}}{i - \frac{2}{\alpha}} \HGF{2}{1}{i+1,i-\frac{2}{\alpha}; i-\frac{2}{\alpha}+1; - s d_0^{-\alpha}} \, .
\end{align}
\noindent where, $\HGF{2}{1}{a,b;c;z}$ is the Gauss hypergeometric function.
\end{theorem}
\begin{IEEEproof}
See Appendix~\ref{app:outageproof}. \hfill \IEEEQEDhere
\end{IEEEproof}
\par From \eqref{eq:pout}, we can see that the uplink outage probability depends on $R$. In the following proposition, we show the effect of $R$ on the outage probability for uplink transmission.
\begin{proposition}
\label{impactRul}
In \eqref{eq:pout}, $\poutul(R)$ is a monotonically decreasing function of the coverage radius $R$, i.e., $\poutul(R)$ monotonically decreases with $R$ and converges to $0$ as $R \to \infty$.
\end{proposition}
\begin{IEEEproof}
Here, it is sufficient to show that $\frac{d}{dR}\poutul(R) \leq 0$. Using the Leibnitz's rule of differentiation under integration in \eqref{eq:pout}, we have
\begin{align}
\label{eq:diffcov}
\nonumber \frac{d}{dR}\poutul(R) & = -2\pi \lambda_b \poutul(R)\\
\nonumber & \hspace{1 cm} \times \, \frac{d}{dR} \int_{0}^{R}r \sum_{m=0}^{M-1}\frac{(-1)^m}{m! \, l^m(r)} \mathcal{L}_{I}^{(m)}\Big(\frac{\widehat{\gamma}}{l(r)}\Big) \, dr\\
\nonumber & = -2\pi \lambda_b \poutul(R)\\ & \hspace{0.5 cm} \times \underbrace{\sum_{m=0}^{M-1} \frac{(-1)^m \, R^{m\alpha+1}}{m!}\mathcal{L}_{I}^{(m)}( \widehat{\gamma} R^{\alpha} )}_{\, \Define \varepsilon(R)} \, .
\end{align}
\indent Note that, from \eqref{eq:diffcov}, we can show $\frac{d}{dR}\poutul(R) \leq 0$, if and only if $\varepsilon(R) \geq 0$ for all $R$. Now, from \eqref{eq:ulltderiv}, we observe that $\mathcal{L}_{I}^{(m)}(\widehat{\gamma} R^{\alpha})$ depends both on $F_1^{(m)}(\widehat{\gamma} R^{\alpha})$ and $F_2^{(m)}(\widehat{\gamma} R^{\alpha})$. From \eqref{eq:partderiv1}, we have
\begin{align}
\label{eq:partderiv11}
\nonumber F_1^{(m)}(\widehat{\gamma} R^{\alpha}) & = \pi \, \lambda_d \, (-1)^m \, F_1(\widehat{\gamma} R^{\alpha})\\
& \hspace{1.5 cm} \times \, \left[ \frac{(\widehat{\gamma} \, d_0^{-\alpha})}{(1 + R^{\alpha} \widehat{\gamma} d_0^{-\alpha})^2} \, + \cdots \right]\, .
\end{align}
\noindent Here in \eqref{eq:partderiv11}, the terms within the bracket are positive and decrease with $R$. Clearly, as $R \to \infty$, $F_1^{(m)}(\widehat{\gamma} R^{\alpha}) \to 0$. Similarly, from \eqref{eq:partderiv2} it can be shown that
\begin{align}
\label{eq:partderiv13}
\nonumber F_2^{(m)}(\widehat{\gamma} R^{\alpha}) & = (-1)^m \, F_2(\widehat{\gamma} R^{\alpha}) \, \pi \lambda_d\\
& \hspace{0.5 cm} \times \, \left[\int_{d_0^2}^{\infty} \frac{(\widehat{\gamma} z^{-\alpha/2})}{(1 + R^{\alpha} \widehat{\gamma} z^{-\alpha/2} )^2} du \, + \cdots \right]\, .
\end{align}
\noindent In \eqref{eq:partderiv13}, the terms within bracket (see the last line) are similarly also positive and decreasing with $R$. Therefore, $F_2^{(m)}(\widehat{\gamma} R^{\alpha}) \to 0$ as $R \to \infty$. From \eqref{eq:ulltderiv} and \eqref{eq:diffcov} it is therefore clear that $\varepsilon(R) \geq 0$ (since $F_1 (\widehat{\gamma} R^{\alpha}) \geq 0$ and $F_2(\widehat{\gamma} R^{\alpha}) \geq 0$), $\forall \,\, R$. \hfill \IEEEQEDhere
\end{IEEEproof}
The result of Proposition~\ref{impactRul} can be intuitively explained as follows. In the uplink, transmission is considered successful as long as at least one of the connected APs can successfully receive the offloaded data transmission. The number of APs connected to a user increases with $R$. Therefore, the probability of successful uplink transmission also increases with $R$. Hence, $\poutul(R)$ decreases monotonically as $R$ increases.
\subsection{Outage Probability for Downlink Transmission}
Once computation of the offloaded data is complete, the processed data is to be transmitted back to the respective users. For this downlink transmission, we consider MRT at each AP connected to the user, and these beamformed data are simultaneously transmitted by all the associated APs to the user. Therefore, the overall received signal at the user becomes the sum of signals received from all the APs. Hence, in the downlink, the user can be in outage if the combined SIR of this total received signal is below a threshold value, i.e.,
\begin{align}
\label{eq:outdl}
\poutdl & \Define \Prb[\text{SIR}_{0,\text{dl}} < \widetilde{\gamma}] \, ,
\end{align}
\noindent where $\text{SIR}_{0,\text{dl}}$ is defined in \eqref{eq:sirdl} and $\widetilde{\gamma}$ is the threshold SIR for downlink transmission. Note that in large wireless networks, it has been shown that the Gamma distribution can provide a tight reasonable approximation of the interference power distribution \cite{Ganti2}. Later, in \cite{Bai,Tanbourgi}, a Gamma approximation of the interference distribution have also been used. Motivated by these findings, we also characterize the total downlink interference power as a Gamma random variable and derive its shaping and scaling parameters in the following proposition. 
\begin{proposition}
\label{intfcharac}
The total interference power, $I_{\text{dl}}$, in \eqref{eq:interfdl} can be approximated as a Gamma distributed random variable with the shaping and scaling parameters, $\zeta(R)$ and $\eta$ respectively, given by
\begin{align}
\label{eq:rvparam1}
\zeta(R) & = \alpha \, \frac{\alpha - 1}{2(\alpha - 2)^2}\lambda_b \, \lambda_d \, \pi^2 \, R^2 \, d_0^2 \, \\
\label{eq:rvparam2}
\eta & = \frac{\alpha - 2}{\alpha - 1} 2 d_0^{-\alpha} \, .
\end{align}
\end{proposition}
\begin{IEEEproof}
The parameters $\zeta(R)$ and $\eta$ of Gamma approximation of the total interference power satisfy the following relations: $\Eb{I_{\text{dl}}} = \zeta(R) \eta$ and $\V{I_{\text{dl}}} = \zeta(R) \eta^2$ \cite{Bai}. Using Campbell' Theorem \cite{Haenggi}, we have
\begin{align}
\label{eq:meanI}
\nonumber \Eb{I_{\text{dl}}} & = 2\lambda_d \pi^2 R^2 \lambda_b \Eb{\widetilde{g}_{kq}} \int_{0}^{\infty} l(r) \, r \, dr\\
& = \frac{\lambda_b \, \lambda_d \, \pi^2 \alpha \, R^2}{\alpha - 2} d_0^{2-\alpha} \, , \hspace{0.3 cm} \text{and}  \\
\label{eq:varI}
\nonumber \V{I_{\text{dl}}} & = 2\pi^2 \lambda_b \lambda_d R^2 \Eb{\widetilde{g}_{kq}^2} \int_{0}^{\infty} l^2(r) \, r \, dr \\
& = \frac{2 \alpha \pi^2 \lambda_b \lambda_d R^2}{\alpha-1} d_0^{2-2\alpha} \, .
\end{align}
\noindent where $\widetilde{g}_{kq}$ is i.i.d. exponential random variable, i.e., $\widetilde{g}_{kq} \sim \text{Exp}(1)$. Using \eqref{eq:meanI} and \eqref{eq:varI}, we obtain \eqref{eq:rvparam1} and \eqref{eq:rvparam2}, respectively. \hfill \IEEEQEDhere
\end{IEEEproof}
\begin{theorem}
\label{trm:outagedl} 
The downlink communication outage probability for the $0$-th user is given by
\begin{align}
\label{eq:poutdl}
\poutdl(R) & \mbox{\scriptsize $ {\gleq \atop {k_0 = \lceil \zeta(R) \rceil}}$ } \sum\limits_{m=0}^{k_0-1}\frac{(-1)^m (\widetilde{\gamma} \eta)^{-m}}{m!}\mathcal{L}_P^{(m)}\left( \frac{1}{\widetilde{\gamma} \eta}\right) \, ,
\end{align}
\noindent where $\mathcal{L}_P(s) \Define \Eb{e^{-sP}}$ is the Laplace transform of the total received signal power, $P \Define \sum_{k \in \bm C(0, R) } g_{k0} \ell(x_{k0})$. $\mathcal{L}_P^{(m)}(s)$ is the $m^{\text{th}}$ order derivative of $\mathcal{L}_P(s)$ and is given by
\begin{align}
\label{eq:lapdevm}
\mathcal{L}_P^{(m)}(s) & = \left \{ \begin{array}{ll}
e^{-2\pi \lambda_b \varrho(s)} & m =0\\
-2 \pi \lambda_b \sum\limits_{i=0}^{m-1} {m-1 \choose i}\mathcal{L}_P^{(i)}(s) \, \varrho^{(m-i)}(s) & m > 0
\end{array} \right . \, .
\end{align}
\noindent Here, $\varrho(s)$ is given by

\begin{align}
\label{eq:varho}
\nonumber \varrho(s) & = \frac{1}{2}{R^2 \left (1 - \frac{1}{(1+sR^{-\alpha})^M} \right)}\\
\nonumber & + \frac{1}{2}\E \left[ (s \, g)^{\frac{2}{\alpha}} \left\{ \Gamma \left(1-\frac{2}{\alpha}, s g R^{-\alpha} \right) \right. \right.\\
& \left. \left. \hspace{2.5 cm} - \, \Gamma \left(1-\frac{2}{\alpha}, s g d_0^{-\alpha} \right)\right\} \right] \, ,
\end{align}

\noindent where $g \sim \Gamma(M,1)$ distributed, and $\varrho^{(m)}(s)$ ($m > 0$) is given by
\begin{align}
\label{eq:paramdev}
\nonumber \varrho^{(m)}(s) & = (-1)^{m-1} \frac{1}{2}  d_0^{2-m \alpha} \frac{\Gamma(M+m)}{\Gamma(M) (1+sd_0^{-\alpha})^{M+m}}  \\
\nonumber & + \, (-1)^{m-1} \frac{1}{\alpha} s^{\frac{2}{\alpha} - m} \E \left[ g^{\frac{2}{\alpha}} \left\{ \Gamma \left(m-\frac{2}{\alpha}, s g R^{-\alpha} \right) \right. \right. \\
&  \hspace{2 cm} \left. \left. - \,\,  \Gamma \left(m-\frac{2}{\alpha}, s g d_0^{-\alpha} \right)  \right\} \right] \, .
\end{align}
\end{theorem}
\begin{IEEEproof}
See Appendix~\ref{app:outageproof2}. \hfill \IEEEQEDhere
\end{IEEEproof}
\begin{remark}
\label{impactRdl}
From \eqref{eq:poutdl}, we can see that the expression on the right hand side (R.H.S.) is a bound on the downlink outage probability, $\poutdl(R)$, and the equality will hold if and only if $\zeta(R)$ is an integer. Furthermore, we observe that this bound on the R.H.S. of \eqref{eq:poutdl} is a weighted sum of $\varrho(\frac{1}{\widetilde{\gamma} \eta})$ and its derivatives. From \eqref{eq:paramdev}, we can write $\chi_m(\frac{1}{\widetilde{\gamma} \eta}) = (-1)^{m-1} \varrho^{(m)}(\frac{1}{\widetilde{\gamma} \eta})$ is a positive quantity and increases with $R$. Therefore, substituting $\varrho^{(m)}(\frac{1}{\widetilde{\gamma} \eta})$ with $(-1)^{m-1} \chi_m(\frac{1}{\widetilde{\gamma} \eta})$ in \eqref{eq:lapdevm}, we can rewrite $\mathcal{L}_P^{(m)}(\frac{1}{\widetilde{\gamma} \eta})$ as a positive sum of $\chi_i(s)$, $i = 1,2, \ldots, m$ and $\varrho(\frac{1}{\widetilde{\gamma} \eta})$, multiplied with $(-1)^{m}$. Using this in \eqref{eq:poutdl} we have $(-1)^{m}\mathcal{L}_P^{(m)}(\frac{1}{\widetilde{\gamma} \eta}) > 0$. In other words, each term on the R.H.S. of \eqref{eq:poutdl} is positive and increases with $R$ (since both $\chi_m(\frac{1}{\widetilde{\gamma} \eta})$ and $\varrho(\frac{1}{\widetilde{\gamma} \eta})$ increases with $R$). We also note that the number of terms on the R.H.S. of \eqref{eq:poutdl} increases with $k_0$, which increases with $\zeta(R)$, an increasing function of $R$ (see \eqref{eq:rvparam1}). In short, the bound on the R.H.S. increases with $R$ and therefore $\poutdl(R)$ is expected to increase as $R$ increases. This can also be intuitively explained from the fact that with increasing $R$, the number of users a single AP beamforms to will increase, and therefore the overall interference from each AP will also increase. This increase in the total received interference power degrades the probability of successful transmission, i.e., in other words, the downlink outage probability increases with $R$.
\end{remark}
\section{Successful Computation Probability Analysis}\label{sec:scp}
In the proposed task offloading model in Section~II-B, it is clear that each MEC server (AP) is connected to all the users requesting edge computing resources within a radius $R$. With the distribution of such users following a homogeneous PPP, the task arrival rate can also be modelled as a Poisson process. Since a user chooses to process the task at the CS with a fixed probability $\vartheta$, the task arrival rate at both the CS and at the MEC servers can be modelled as Poisson processes. For analytical tractability, we assume that all MEC servers are of equal computation capacity. In the following, we now first derive the task arrival rates at the CS and at the MEC servers and then derive the successful computation probability (SCP) expression for the proposed task computation model.
%
%
%
%
\begin{proposition}
\label{taskarrive}
The task arrival rates at the CS and a MEC server are respectively given by
\begin{align}
\label{eq:lamcs}		
		\Lc 
		&	=	
			\lambda_d |\mathcal{A}| \vartheta 
			\left( 1-  \poutul(R) \right)
		\\
		\label{eq:lammec}
		\Lm
		& = 
			\frac{\lambda_d}{\lambda_b} (1-\vartheta) \left(1-e^{-\lambda_b \pi R^2}\right) \left( 1-  \poutul(R) \right) 
\end{align}
\noindent where $|\mathcal{A}|$ denotes the total network area and $\poutul(R)$ is presented in \eqref{eq:pout}.
\end{proposition}
\begin{IEEEproof}
In the proposed task offloading model, a CS or MEC server executes 
a offloaded task, which is successfully received by at least one of MEC servers connected to the user offloading the task. 
Hence, in the network, the total rate of task arrivals at the computing devices (i.e., CS and MEC servers)  is  $\lambda_d |\mathcal{A}| \{1 - \poutul(R)\}$.
Since $\vartheta$ portion of the total task arrivals at computing devices is assigned to the CS, 
the task arrival rate at the CS becomes $\Lc = \vartheta \lambda_d |\mathcal{A}| \left(1 - \poutul(R)\right)$.
On the other hand, a MEC server receives offloaded tasks from all users in its coverage, i.e., a circle with radius $R$, and the rate of offloading tasks to a MEC server becomes  
\begin{align}	
	\lambda_\text{o} = (1-\vartheta) \lambda_d \pi R^2 \left( 1- \poutul(R) \right) \,.
\end{align}
\indent From Section~II-B.2, we note that only the MEC server with minimum computation load processes the task. Since all the MEC servers have equal computation capability, the probability that a connected MEC server would be selected for processing the task is $\frac{1}{N^{\prime}+1}$, where $N^{\prime}$ is the number of connected MEC servers, besides the tagged server. Since $N^{\prime}$ is a Poisson random variable with mean $\lambda_b \pi R^2$, the probability that the tagged MEC server becomes the executing server of the task, $p_{m, \text{min}}$, is given by
\begin{align}
\label{eq:mecprob1}
\nonumber p_{m, \text{min}} & = \Eb{\frac{1}{1+ N^{\prime}}} \, = \, \sum\limits_{k=0}^{\infty} e^{-\lambda_b \pi R^2} \frac{(\lambda_b \pi R^2)^k}{k!}\frac{1}{k+1}\\
& = \sum\limits_{n=1}^{\infty} e^{-\lambda_b \pi R^2} \frac{(\lambda_b \pi R^2)^n}{n! \lambda_b \pi R^2} \, = \, \frac{(1-e^{-\lambda_b \pi R^2}) }{\lambda_b \pi R^2} \, .
\end{align}
%
%
%
Therefore, the task arrival rate at a MEC server for execution becomes $\Lm = \lambda_\text{o} p_{m, \text{min}}$, which is given in \eqref{eq:lammec}.\hfill \IEEEQEDhere
\end{IEEEproof}
%

%
%
%
\subsection{Computation Latency Analysis} \label{latmodel}

The successful computation probability (SCP) is defined as the probability of the event that task processing is completed within a target computation time. To compute SCP, we use queueing system concept to model $\Tcs$ and $\Tmec$ separately for different users. In the following, we derive the expression for SCP for a given target computation latency.
{\begin{theorem}\label{trm:scp}
{Using the definition of SCP from \eqref{eq:scpdef}, we have}
{\begin{align}
\label{eq:compcdf}
p_{\text{comp}}(R, \vartheta, t) 
	& = \vartheta \Prb[T_{\text{c}} \leq t] + (1-\vartheta) \Prb[\Tmec \leq t],
\end{align}
where $\Prb[T_{\text{c}} \leq t]$ and $\Prb[\Tmec \leq t]$ are respectively given by 
	\begin{align}
	\label{eq:tcsi}
		\Prb[T_{\text{c}} \leq t] 
		= 
			\int_{0}^{t} 
			\mathcal{L}_{\Tcs}^{-1}
			\left[
				\frac{(1-\rho_{\text{c}})s\sum\limits_{i=1}^{\mathcal{I}}\frac{p_i \mu_{\text{c},i}}
				{s+\mu_{\text{c},i}}}{s - \lambda_{\text{c}} + \lambda_{\text{c}} 
				\sum_{i=1}^{\mathcal{I}}\frac{p_i \mu_{\text{c},i}}{s+\mu_{\text{c},i}} }
			\right]
		du
		\, ,
	\end{align}}
	{\begin{align}
	\label{eq:mecPnew}
	\nonumber 
	\Prb[\Tmec \leq t] 
		& = 
			\sum\limits_{i=1}^{\I} 
				p_i \sum\limits_{n=1}^{\infty}
				\frac{(\lambda_b \pi R^2)^n}{n!}e^{-\lambda_b \pi R^2}\\
		\nonumber 		& \times \,\,
		\sum\limits_{v=0}^{\infty} 
		\left[\left( 
			\sum\limits_{i=1}^{\I} \frac{\epsilon_i \omega_i^{v}}{1 - \omega_i} \right)^n 
			- 
			\left( \sum\limits_{i=1}^{\I} \frac{\epsilon_i \omega_i^{v+1}}{1 - \omega_i} 
		\right)^n \right]\\
		& \hspace{0.2 cm}\times \, 
		\int_{0}^{t}
		 	\mathcal{L}^{-1} 
			\left[ 
				\left( 
					\sum\limits_{i=1}^{\I} 
					\frac{p_i \mu_{\text{m},i}}{s+\mu_{\text{m},i}} 
				\right)^{v+1} 
			\right]
		 du
		\, ,
	\end{align}}
%
%
%
%
%
\hspace{-0.3 cm} {\noindent 
for the inverse Laplace transform of $X$ $\mathcal{L}_{X}^{-1}(\cdot)$,
$\rho_{\text{c}} \Define \frac{\lambda_{\text{c}}}{\mu_{\text{c}}}$, 
and $\mu_{\text{c}} \Define \Big(\sum_{i=1}^{\I}\frac{p_i}{\mu_{\text{c},i}} \Big)^{-1}$.}
\hspace{-0.3 cm} {\noindent In \eqref{eq:mecPnew}, $\omega_i$ ($i = 1,2,\ldots, \I$) are the solutions of the following equation for $\omega$,}
{\begin{align}
\label{eq:mec131}
\nonumber \sum\limits_{l=1}^{\I}p_l \mu_{\text{m},l} \omega \left\{\prod\limits_{k=1, k\neq l}^{\I}( \omega(\mu_{\text{m},k} + \Lm) - \Lm ) \right\} & \\
 & \hspace{-2.5 cm} = \prod\limits_{q=1}^{\I}(\omega(\mu_{\text{m},q} + \Lm) - \Lm ) \, ,\tag{37}
\end{align}}
\hspace{-0.3 cm} {\noindent and $\epsilon_i$, $\forall i = 1, 2, \ldots, \I$, are obtained from comparing the coefficients of different powers of $z$ in the following equality}
{\begin{align}
\label{eq:mec132}
\frac{\sum\limits_{l=1}^{\I}p_l \mu_{\text{m},l} \prod\limits_{k=1, k\neq l}^{\I}(\mu_{\text{m},k} + \Lm - \Lm z)}{\sum\limits_{q=1}^{\I}\epsilon_q \prod\limits_{r=1,r\neq q}^{\I}(1-\omega_r z)} & = \frac{1}{(1-\rho_{\text{m}})(1-z)} \, .\tag{38}
\end{align}}
\hspace{-0.3 cm} {\noindent Here, $\rho_{\text{m}} \Define \frac{\lambda_{\text{m}}}{\mu_{\text{m}}}$, and $\mu_{\text{m}} \Define \Big(\sum_{i=1}^{\I}\frac{p_i}{\mu_{\text{m},i}} \Big)^{-1}$.}
\end{theorem}}

\begin{IEEEproof}
See Appendix~\ref{app:scpmg1}. 
\end{IEEEproof}
{\begin{corollary}
\label{mm1new}
For $\I = 1$, the SCP is given by \eqref{eq:compcdf}, where $\Prb[\Tcs \leq t] = 1 - e^{-\mu_{\text{c}} t + \Lc t}$, and $\Prb[\Tmec \leq t]$ is given by
{\begin{align}
\label{eq:mm1qnew}
\Prb[\Tmec \leq t] & = \sum\limits_{n=1}^{\infty}\frac{(\lambda_b \pi R^2)^n}{n!}e^{-\lambda_b \pi R^2} \left( 1 - e^{-\mu_{\text{m}} t(1-\rho_{\text{m}}^n)} \right) \, .
\end{align}}
{\noindent for $\mu_{\text{m},1} = \mu_{\text{m}}$ and $\mu_{\text{c},1} = \mu_{\text{c}}$.}
\end{corollary}}
\begin{IEEEproof}
{With $\mathcal{I} = 1$, the queueing system at the CS or at a MEC server becomes M/M/1. Clearly, for the service rate of $\mu_{\text{c}}$, the distribution of computation latency at the CS is $f_{T_{\text{c}}}(u) = (\mu_{\text{c}} - \Lc)e^{-(\mu_{\text{c}} - \Lc)u}$ ($u \geq 0$), so $\Prb[\Tcs \leq t] = 1 - e^{-\mu_{\text{c}} t + \Lc t}$. In addition, with $\I = 1$, $N_{\text{m},k}$ in \eqref{eq:latmecdefnew} is a geometric random variable with parameter $1-\rho_{\text{m}}$ \cite{Kleinrock}. Therefore, from \eqref{eq:tqcdf}, we have $\Prb[\bar{N}_{\text{m}}(n) = v] = (1 - \rho_{\text{m}}^n) \rho_{\text{m}}^{n v} $. For a given $\bar{N}_{\text{m}}(n)$, from \eqref{eq:tmecLT}, we have $f_{\Tmec | \bar{N}_{\text{m}}(n) = v}(x) = \frac{ \mu_{\text{m}}^{v+1} x^v e^{-\mu_{\text{m}} x}}{\Gamma(v+1)}$ ($x \geq 0$). Using this in \eqref{eq:mec1}, we have} 
{\begin{align}
\label{eq:mec12}
\Prb[\Tmec \leq t | N = n] & = (1 - \rho_{\text{m}}^n)   \, \underbrace{\sum\limits_{v=0}^{\infty}\rho_{\text{m}}^{n v}  F(v+1, \mu_{\text{m}} t )}_{\Define V_{\infty} }  \, .
\end{align}}
{\noindent where $F(r+1,z) = \frac{\gamma(r+1, z)}{\Gamma(r+1)}$ ($r \geq 0$ and $z > 0$), and $\gamma(s,x) \Define \int\limits_{0}^{x} t^{s-1} e^{-t} \, dt$ denotes the lower incomplete Gamma function. Using integration by parts in the definition of $\gamma(s,x)$, we can show that $F(r+1, z) = F(r,z) - \frac{z^r}{r!} e^{-z}$, $\forall \, r > 0$, i.e., $F(r+1, z) = F(1,z) - \sum\limits_{a=1}^{r} \frac{z^a}{a!} e^{-z} $, and $F(1,z) = 1 - e^{-z}$. Hence $V_{\infty}$ becomes}
{\begin{align}
\label{eq:interm1}
\nonumber V_{\infty} & = \sum\limits_{v=0}^{\infty} F(v+1,\mu_{\text{m}} t)\rho_{\text{m}}^{n v}\\
\nonumber & = \sum\limits_{v=0}^{\infty}\rho_{\text{m}}^{n v} \left(F(1,\mu_{\text{m}} t) - \sum\limits_{a=1}^{v} \frac{(\mu_{\text{m}} t)^a}{a!}e^{-\mu_{\text{m}} t} \right)  \\
\nonumber &  = \sum\limits_{v=0}^{\infty}\rho_{\text{m}}^{n v} F(1,\mu_{\text{m}} t) - e^{-\mu_{\text{m}} t}\sum\limits_{w=1}^{\infty} \frac{(\mu_{\text{m}} t)^w \rho_{\text{m}}^{n w} }{w!}\sum\limits_{r=0}^{\infty}\rho_{\text{m}}^{n r}\\
&  = \frac{1 - e^{-\mu_{\text{m}} t(1-\rho_{\text{m}}^n)}}{1 - \rho_{\text{m}}^n} \, .
\end{align}}
{\noindent Substituting \eqref{eq:interm1} in \eqref{eq:mec12}, we obtain \eqref{eq:mm1qnew}.}
\end{IEEEproof}

\subsection{Impact of System Parameters on SCP} \label{impactRtheta}

From the analysis in Section~\ref{latmodel}, we observe that it is difficult to directly characterize the impacts of $R$ and $\vartheta$ on the successful computation probability (SCP), $\pco(R, \vartheta, t)$, for a general case of $\I$. Therefore, in this section, we first analyze their impact for the special case used in Corollary~\ref{mm1new} (i.e., $\I = 1$) and then extend it for higher values of $\mathcal{I}$.
\subsubsection{Impact of $\vartheta$}
In the following, we show that for any given $R$ and $t$, the SCP is maximized for some $\vartheta \in [0,1]$.
{\begin{proposition}
\label{impacttheta}
The $\pco(R, \vartheta, t)$ increases with $\vartheta$, for all $\vartheta \leq \vartheta_0$ and decreases with $\vartheta$ for $\vartheta > \vartheta_0$, where $\vartheta = \vartheta_0$ satisfies $\frac{\partial}{\partial \vartheta}\pco(R, \vartheta, t)\Big|_{\vartheta = \vartheta_0} = 0$.
\end{proposition}}
{\begin{IEEEproof}
Here, it is sufficient to show that $\pco(R, \vartheta, t)$ is maximum for $\vartheta = \vartheta_0$ for a given $R$ and $t$. To show this, we compute the first and second order derivatives of $\pco(R, \vartheta, t)$ with respect to $\vartheta$, as given below
\begin{align}
\label{eq:scpdiff}
\nonumber \frac{\partial}{\partial \vartheta} & \pco(R, \vartheta, t)  = 1 - a_1 e^{k_1\vartheta t}(1 + k_1\vartheta t)\\
\nonumber &  \quad \quad \quad  \,\, -\sum\limits_{n=1}^{\infty}k_2(n) \left(1 - a_2^{1 - a_3^n(1-\vartheta)^n}\right) \\
&   \quad   \,\, - (1-\vartheta)\sum\limits_{n=1}^{\infty}k_2(n) a_2^{1 - a_3^n(1-\vartheta)^n} n a_3^n \ln (a_2^{-1})\\
\nonumber \frac{\partial^2} {\partial \vartheta^2}  & \pco(R, \vartheta, t)  = -a_1 k_1 t(2+k_1 \vartheta t)e^{k_1 \vartheta t}\\
\nonumber &\quad -\sum\limits_{n=1}^{\infty} n^2 k_2(n)a_3^n \ln(a_2^{-1})a_2^{1 - a_3^n(1-\vartheta)^n} (1-\vartheta)^{n-1}\\
 & \quad \quad\quad  \times \, \left[1 +(1-\vartheta)^n a_3^n \ln(a_2^{-1})\right]
\label{eq:scpdiff2}
\end{align}
\noindent where $a_1 = e^{-\mu_c t}$, $a_2 = e^{-\mu_m t}$, $a_3 = \frac{\Lm}{\mu_m(1-\vartheta)}$, $k_1 = \lambda_d |\mathcal{A}|\{1-\poutul(R)\}$, and $k_2(n) = \frac{(\lambda_b \pi R^2)^n}{n!}e^{-\lambda_b \pi R^2}$. Equating $\frac{\partial}{\partial \vartheta}\pco(R, \vartheta, t) = 0$, we obtain $\vartheta = \vartheta_0$, and from \eqref{eq:scpdiff2} we have $\frac{\partial^2}{\partial \vartheta^2}\pco(R, \vartheta, t) < 0$. This shows that $\vartheta = \vartheta_0$ maximizes $\pco(R, \vartheta, t)$ for a given value of $R$ and $t$.  \hfill \IEEEQEDhere
\end{IEEEproof}}
\par The above result can be explained as follows. When $\vartheta$ is sufficiently small, the task is processed at a MEC server with high probability. Since the task arrival rate at the MEC servers, $\Lm$, decreases with $\vartheta$, $\Prb[\Tmec \leq t]$ increases with $\vartheta$ and therefore, $\pco(R, \vartheta, t)$ increases with $\vartheta$. On the other hand, when $\vartheta \to 1$, task processing occurs at the CS with high probability. Since the task arrival rate at the CS, $\Lc$, increases with $\vartheta$, in this regime, $\pco(R, \vartheta, t)$ decreases with $\vartheta$. This shows that there exists some $\vartheta \in [0,1]$ for a given $R$ and $t$ that maximizes $\pco(R, \vartheta, t)$. Note that the above discussion holds true, irrespective of the value of $\I$. Therefore, the result in Proposition~\ref{impacttheta} is also applicable for $\I > 1$ scenario.

\subsubsection{Impact of $R$} 
To study the impact of $R$ on the SCP, we consider the following scenarios: (a) when both $R$ and $\vartheta$ are small; (b) $R$ is small and $\vartheta \to 1$; and (c) when $R$ is sufficiently large. Note that the number of connected MEC servers increases with $R$. Therefore, in the first scenario (i.e. $\vartheta \to 0$ and small $R$), the probability of obtaining a smaller value of the minimum queue length increases with $R$. In other words, in this regime, $\Prb[\Tmec \leq t]$ (i.e. $\pco(R, \vartheta, t)$) increases with $R$. On the other hand, when $\vartheta \to 1$, the task processing occurs at the CS with high probability and $\Lc$ increases with $R$ (since $\poutul(R)$ decreases with $R$). Clearly, in this regime, the SCP decreases with $R$. However, when $R$ is sufficiently large, $\poutul(R) \to 0$ and therefore in this regime, both $\Lc$ and $\Lm$ converge to constant values. Consequently, the SCP, $\pco(R, \vartheta, t)$ also saturates to a fixed value in this regime, irrespective of the value of $\vartheta$.

\section{Numerical Results}\label{sec:results1}

In this section, we analyze the impact of various system parameters (e.g., $R$, $\vartheta$, $\lambda_b$) on the proposed performance metrics, i.e., $p_{\text{scmp}}(R)$, $p_{\text{comp}}(R, \vartheta, t)$ and $p_{\text{secp}}(R, \vartheta, t)$.

\subsection{Successful Communication Probability}
\begin{figure}[t!]
    \begin{center}   
    { 
\hspace{0.1 cm}
	 \includegraphics[width=1.00\columnwidth]{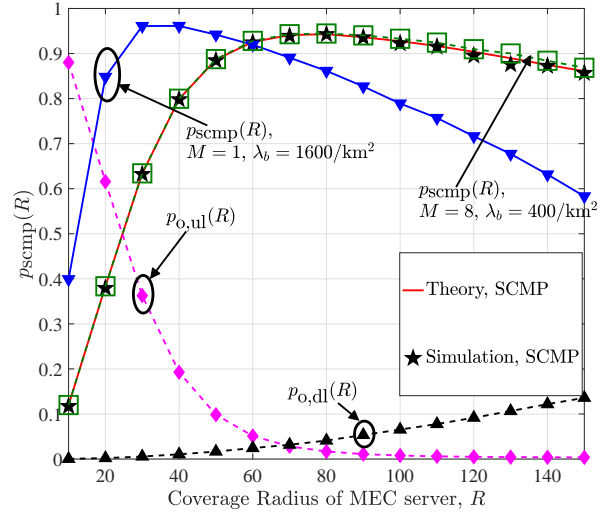}
    }
    \end{center}
    \caption{
    		Successful communication probability, $p_{\text{scmp}}(R)$, as a function of coverage radius, $R$. }
   \label{fig:scmpuldloutage}
\end{figure}

In Fig.~\ref{fig:scmpuldloutage} we plot the successful communication probability, $p_{\text{scmp}}(R)$ (defined in \eqref{eq:scmpdef}) both numerically and through simulation as a function of $R$ for the following fixed parameters: $M = 4$, $\lambda_b = 400$/km\textsuperscript{2}, $\lambda_d = 100$/km\textsuperscript{2}, $\alpha = 3.7$, $d_0 = 1$ m, and $\widehat{\gamma} = \widetilde{\gamma} = 2.62$ dB (i.e., the threshold information rate for uplink and downlink transmissions is $1.5$ bps/Hz). From Fig.~\ref{fig:scmpuldloutage}, it is observed that for small $R$, the SCMP increases with $R$. This is due to the fact that when $R$ is sufficiently small, $\poutdl(R)$ is also small (see the curve with filled triangles in Fig.~\ref{fig:scmpuldloutage}), and therefore SCMP is dominated by $\poutul(R)$ in this regime. Since $\poutul(R)$ decreases rapidly with $R$, in this regime, SCMP increases with $R$. On the other hand, when $R$ is sufficiently large, i.e., when $\poutul(R) \approx 0$ (see Proposition~\ref{impactRul}), the SCMP is dominated by $\poutdl(R)$. Since $\poutdl(R)$ increases with $R$, in this regime, the SCMP begins to decrease with $R$. We also plot the SCMP for the following two scenarios: (a) $M = 1$, $\lambda_b = 1600$/km\textsuperscript{2} (see the curve with blue triangles); and (b) $M = 8$, $\lambda_b = 400$/km\textsuperscript{2} (see the dotted curve with green squares). It is observed that for small values of $M$, the changes in the SCMP is negligible. On the other hand, with the same antenna density (i.e., $M \lambda_b = 1600$/km\textsuperscript{2}), it is observed that in the small $R$ regime, the SCMP is higher for the scenario with larger AP density ($\lambda_b = 1600$/km\textsuperscript{2}), compared to the scenario with smaller AP density ($\lambda_b = 400$/km\textsuperscript{2}). This is due to the fact that in the small $R$ regime, SCMP is dominated by $\poutul(R)$ and $\poutul(R)$ rapidly decreases with AP density, $\lambda_b$ (see \eqref{eq:pout}). On the other hand, in the large $R$ regime, $\poutul(R) \to 0$ and SCMP is dominated by $\poutdl(R)$, which increases with AP density, $\lambda_b$. Therefore, in this regime, SCMP is higher for the scenario with smaller AP density. In other words, for a fixed antenna density, increasing number of APs will improve the probability of successful communication, only when the coverage radius $R$ is sufficiently small.

%
%
%

\begin{figure}[t!]
    \vspace{-0.3 cm}
    \begin{center}   
    { 
	 \includegraphics[width=1.00\columnwidth]{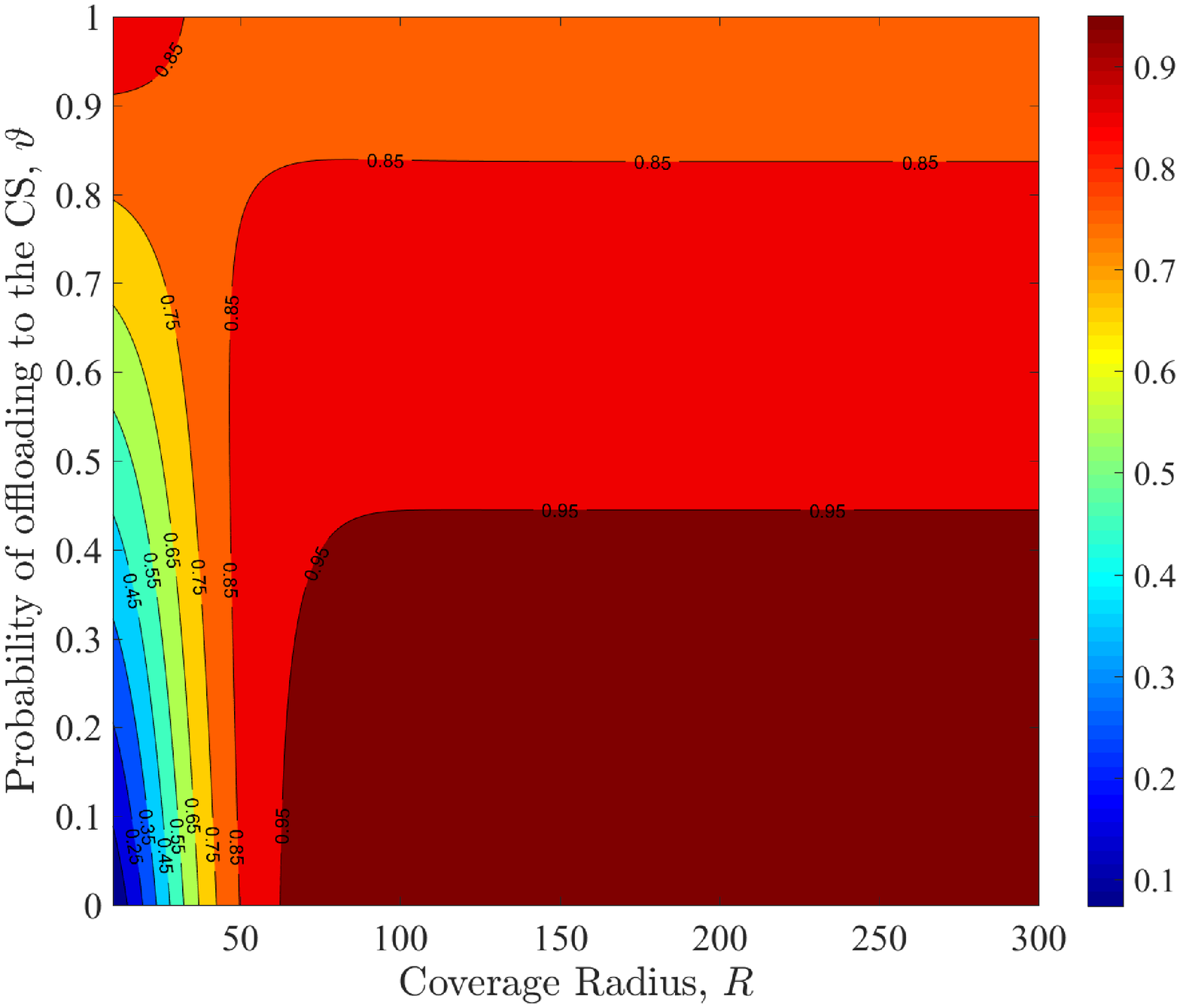}
    }
    \end{center}
    \caption{
    		Successful computation probability, $p_{\text{comp}}(R,\vartheta,t)$, as functions of $(R, \vartheta)$ with $t = 12$ ms, $\mathcal{I}=2$, $p_1 = 0.6$, $p_2 = 0.4$, $M = 4$, $\lambda_b = 400$/km\textsuperscript{2}, and $\lambda_d = 100$/km\textsuperscript{2}.	
		 } 
   \label{fig:scpphlb}
\end{figure}

\begin{figure}[h]
    \vspace{-0.3 cm}
    \begin{center}   
    { 
	 \includegraphics[width=1.00\columnwidth]{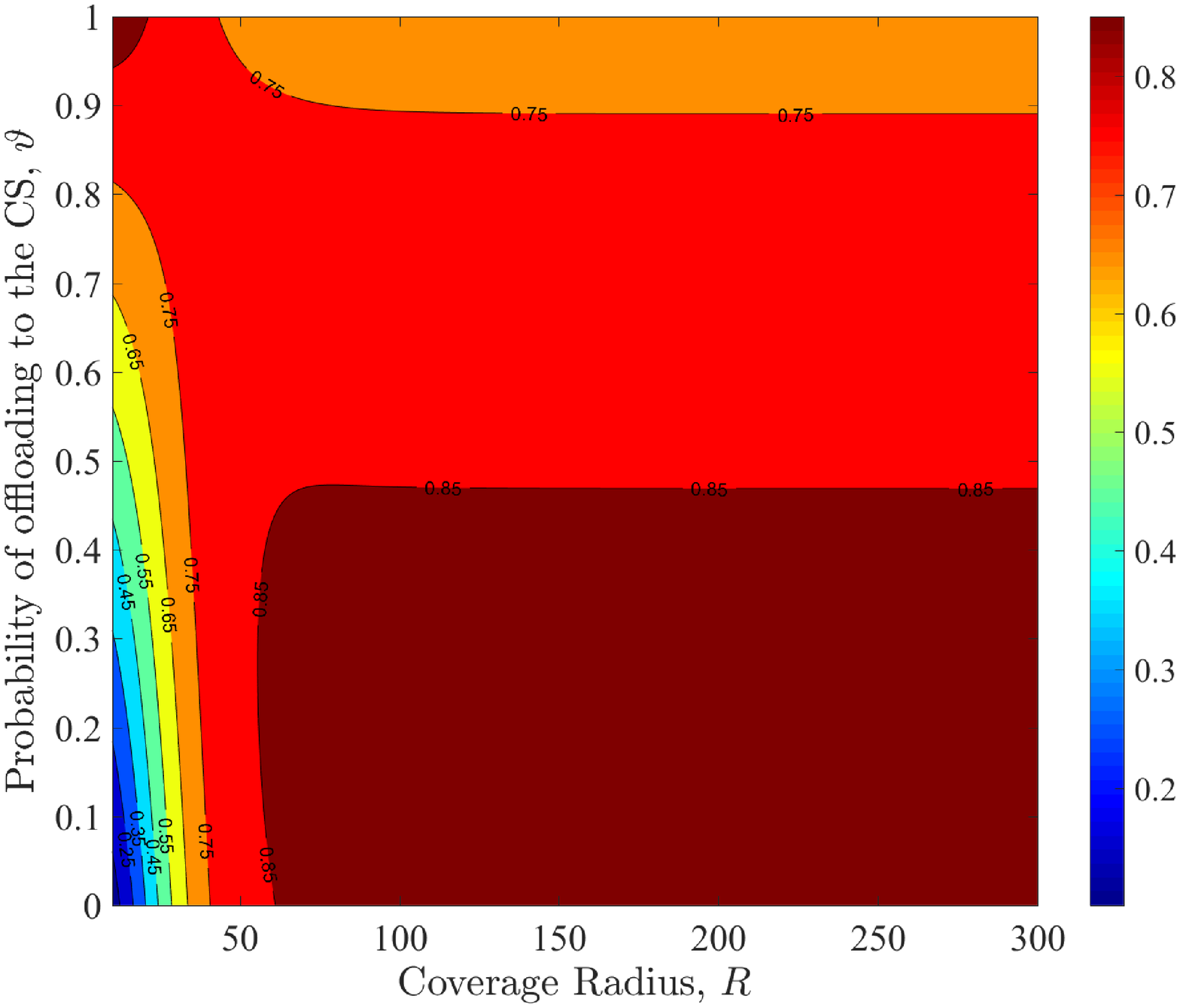}
    }
    \end{center}
    \caption{
    		Successful computation probability, $p_{\text{comp}}(R,\vartheta,t)$, as functions of $(R, \vartheta)$ with $t = 12$ ms, $\mathcal{I}=1$, $M = 4$, $\lambda_b = 400$/km\textsuperscript{2}, and $\lambda_d = 100$/km\textsuperscript{2}.
		 } 
   \label{fig:scpph1new}
\end{figure}

\subsection{Successful Computation Probability}

Next, we analyze the SCP, $p_{\text{comp}}(R, \vartheta, t)$, as functions $R$ and $\vartheta$. The values of the edge computing parameters used for numerical results are given in Table~\ref{table:varMpu}. We plot the contour of $p_{\text{comp}}(R, \vartheta, t)$ with $\I = 2$, for the following two cases: (a) $\I = 2$, $p_1 = 0.6$, $p_2 = 0.4$ (see Fig.~\ref{fig:scpphlb}); and (b) $\I = 2$,$p_1 = 0$, $p_2 = 1$ (i.e., $\I = 1$) (see Fig.~\ref{fig:scpph1new}). In both cases, for a given $R$, it is observed that the SCP is maximized for some $\vartheta \in [0,1]$. Furthermore, we also observe that when both $R$ and $\vartheta$ are small, $\pco(R, \vartheta, t)$ increases with both $R$ and $\vartheta$. On the other hand, when $\vartheta$ is sufficiently large, $\pco(R, \vartheta, t)$ decreases with $R$. This supports our discussion in Section~\ref{impactRtheta}.

\begin{figure}[t!]
    \begin{center}   
    { 
	 \includegraphics[width=0.95\columnwidth]{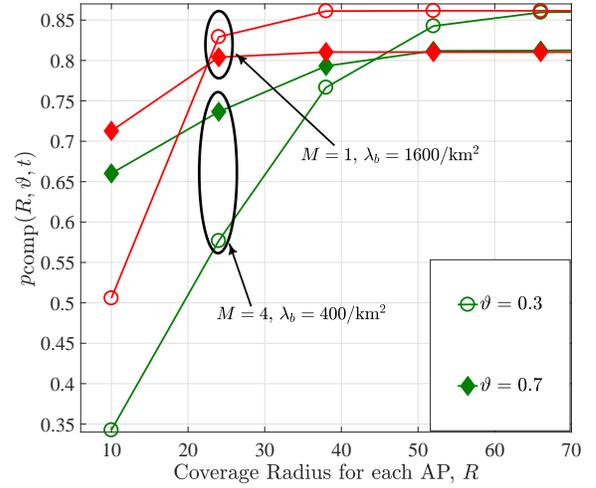}
    }
    \end{center}
    \caption{
    		Successful computation probability, $p_{\text{comp}}(R,\vartheta,t)$, as a function of $R$ with $t = 12$ ms, $\mathcal{I}= 1$, and $\lambda_d = 100$/km\textsuperscript{2} with varying $M$ and $\lambda_b$.
		 } 
   \label{fig:scpR}
\end{figure}

In Fig.~\ref{fig:scpR}, we plot $\pco(R, \vartheta, t)$ as a function of $R$ for $\vartheta = 0.3$ and $\vartheta = 0.7$, for a fixed antenna density $M \lambda_b = 1600$/km\textsuperscript{2} (for $\I = 1$ scenario in Fig.~\ref{fig:scpph1new}). It is also observed that the SCP is higher for smaller $\vartheta$, when $R$ is sufficiently large. {Note that since the CS is connected to all the APs in the CF massive MIMO system, increasing the probability of offloading to the CS would increase the overall task arrival rate at the CS, thereby incurring significant computation latency. On the other hand, when sufficient number of connected MEC servers are available, choosing MEC server for processing (i.e., smaller $\vartheta$) would improve the SCP. This is also the motivation for introducing the edge computing concept in the CF massive MIMO system.} Finally, it is observed that for a fixed antenna density, fixed $\vartheta$, and sufficiently large $R$, the SCP is higher for the scenario with higher AP density (e.g., $\lambda_b = 1600$/km\textsuperscript{2}), compared to the scenario with smaller AP density (e.g., $\lambda_b = 400$/km\textsuperscript{2}).

\begin{figure}[t]
    \begin{center}   
    { 
	 \includegraphics[width=1.00\columnwidth]{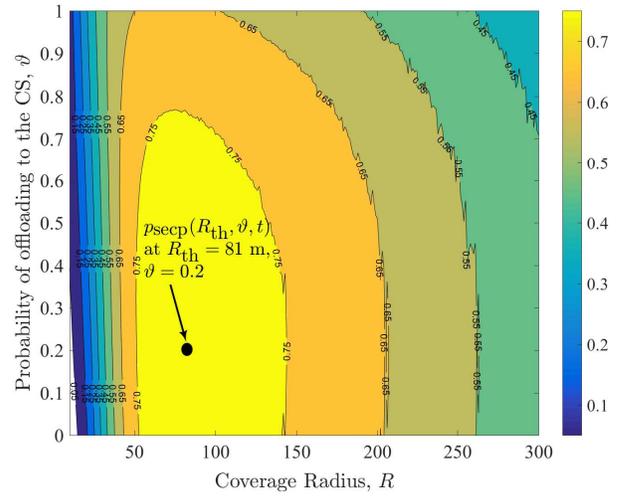}
    }
    \end{center}
    \caption{
    		Successful edge computing probability, $p_{\text{secp}}(R,\vartheta,t)$, as functions of $(R, \vartheta)$ with $t = 12$ ms, $\mathcal{I}=1$, $\lambda_b = 400$/km\textsuperscript{2}, and $\lambda_d = 100$/km\textsuperscript{2}.	
		 }
   \label{fig:secpph10}
\end{figure}

\subsection{Successful Edge Computing Probability}\label{secpres}

Finally, in Fig.~\ref{fig:secpph10}, we combine both the communication and computation performance metrics to evaluate the overall edge computing performance (for the computation scenario in Fig.~\ref{fig:scpph1new}). Here, we plot $p_{\text{secp}}(R, \vartheta, t)$ as functions of $R$ and $\vartheta$. We observe that in the small $R$ regime, the SECP increases with $\vartheta$. This is due to the fact that in this regime, due to insufficient number of connected MEC servers, it is preferable that we process the tasks at the CS for successful edge computing. On the other hand, when $R$ is sufficiently large, the SECP is observed to decrease with $\vartheta$. This is due to the fact that in this regime, with sufficient number of connected MEC servers being available, the probability of successful computation increases for the MEC server processing scenario. Similarly, we also observe that the SECP first increases with $R$ for a given $\vartheta$ and then decreases. The initial increase in SECP with $R$ is due to the fact that for small $R$, the downlink outage probability, $\poutdl(R)$, is negligible, while both the uplink outage probability, $\poutul(R)$, and SCP improve significantly. On the other hand, when $R$ is sufficiently large, $\poutul(R)$ becomes negligible and the SCP saturates to a fixed value. in this regime, the $\poutdl(R)$ increases with $R$ and therefore the SECP begins to decrease. We denote this optimal value of $R$ that maximizes the SECP as $R_{\text{th}}$.{\footnote[8]{{From the plot of SECP in Fig.~\ref{fig:secpph10}, we observe that the SECP is a quasi-concave function of $R$. Therefore, a suboptimal value for $R_{\text{th}}$ can be computed using the bisection searching method \cite{Boyd}. This approach also holds for the SCMP.}}}



 In Table~\ref{table:maxR}, we show the required $R_{\text{th}}$ for different values of $\lambda_b$ and $t$. It is observed that when $\lambda_b$ is fixed, the $R_{\text{th}}$ slowly decreases with $t$. This is due to the fact that for a given value of SCP, the minimum required $R$ decreases with $t$. Since the SCP increases with $t$, any increase in $t$ will therefore result in the SCP reaching its saturation value for even a smaller value of $R$. Consequently, the SECP will be maximized for even a smaller value of $R_{\text{th}}$. Due to the similar reasons, $R_{\text{th}}$ is also observed to decrease with $\lambda_b$.

\begin{table}[t]
\caption{Threshold $R$ and $\vartheta$ that maximizes $p_{\text{secp}}(R, \vartheta, t)$. \label{table:maxR}} 
\begin{center}
\rowcolors{2}
{cyan!15!}{}
\renewcommand{\arraystretch}{1.5}
\begin{tabular}{c|c|c|c|c}
\hline 
 {\bf $M$} & {\bf $\lambda_b$ (in per km\textsuperscript{2})} & {\bf $t$ (in ms)} & {\hspace{0.32cm}}{\bf $R_{\text{th}}$ (in m)} & {$\vartheta$}\\
\hline 
												\hspace{0.15cm}$4$ & \hspace{0.2cm}$400$ 
															& \hspace{0.12cm}$4$ & \hspace{0.2cm}$82$  &{0.19} \\ 
															\hspace{0.15cm}$4$ & \hspace{0.2cm}$400$ 
																		& \hspace{0.12cm}$12$ & \hspace{0.2cm}$81$  &{0.20} \\ 
												\hspace{0.15cm}$1$ & \hspace{0.2cm}$1600$ 
																					& \hspace{0.12cm}$4$ & \hspace{0.2cm}$57$  &{0.21} \\ 
																					\hspace{0.15cm}$1$ & \hspace{0.2cm}$1600$ 
																								& \hspace{0.12cm}$12$ & \hspace{0.2cm}$55$  &{0.25} \\ 
\hline
\end{tabular} 
\end{center}
\end{table}%

\section{Numerical Results on Energy Consumption}\label{sec:results2}

In this section, we analyze the relationship between the SECP and the total energy consumption in edge computing-enabled CF massive MIMO systems. For this, we first formulate the following energy consumption minimization problem:
%
%
\begin{align}
\label{eq:mineconsumption}
\nonumber P : \,\, & \underset{R, \, \vartheta}{\text{minimize}}
& & E(R, \, \vartheta,t)\\
& \text{subject to}
& & p_{\text{secp}}(R,\vartheta,t) \geq \xi 
\end{align}
\noindent where $E(R, \, \vartheta,t)$ denotes the total energy consumption for the target computation latency $t$ and $\xi$ denotes the minimum desired value of the SECP. From our discussion in Section~\ref{secpres}, we note that the feasible range of $(R, \vartheta)$ that satisfies $p_{\text{secp}}(R,\vartheta,t) \geq \xi$ can be obtained from Fig.~\ref{fig:secpph10}. In the following, we first characterize the total energy consumption as functions of $R$ and $\vartheta$, and then analyze the impact of the minimum desired SECP, $\xi$, on $E(R, \, \vartheta,t)$.

\begin{table}[!t]
\caption{Communication Power Consumption Parameters  \label{table:Eparam}} 
\begin{center}
\rowcolors{2}
{cyan!15!}{}
\renewcommand{\arraystretch}{1.5}
\begin{tabular}{l | l | l}
\hline 
 {\bf Parameters} & {\bf Descriptions} & {\bf Values}  \\
\hline 
\hspace{0.15cm}$P_{\text{b}}$ & \hspace{0.2cm}RF chain power at AP
			& \hspace{0.12cm}$0.01$ Watt  \\ 
\hspace{0.15cm}$P_{\text{d}}$ & \hspace{0.2cm}RF chain power at user
			& \hspace{0.12cm}$0.01$ Watt \\ 
\hspace{0.15cm}$P_{\text{osc}}$  & \hspace{0.2cm}Local Oscillator power
			& \hspace{0.12cm}$2$ Watt    \\ 
			\hspace{0.15cm}$P_{\text{cod}}$  & \hspace{0.2cm}Channel coding power
						& \hspace{0.12cm}$0.1$ $\frac{\text{Watt}}{\text{Gbits/sec}}$   \\ 
\hspace{0.15cm}$P_{\text{dec}}$  & \hspace{0.2cm}Channel decoding power
			& \hspace{0.12cm}$0.8$ $\frac{\text{Watt}}{\text{Gbits/sec}}$   \\ 
\hspace{0.15cm}$P_u$ & \hspace{0.2cm}Per-user transmit power
			& \hspace{0.12cm}$0.1$ Watt   \\ 
			\hspace{0.15cm}$P_t$ & \hspace{0.2cm}Per-AP transmit power
						& \hspace{0.12cm}$1.181$ Watt   \\ 
\hspace{0.15cm}$P_{\text{ue}}$  & \hspace{0.2cm}Fixed power at user
			& \hspace{0.12cm}$0.1$ Watt  \\ 
\hspace{0.15cm}$P_{\text{bs}}$  & \hspace{0.2cm}Fixed power at AP
			& \hspace{0.12cm}$5$ Watt  \\ 
\hspace{0.15cm}$\frac{1}{\zeta}$ & \hspace{0.2cm}Power amplifier (PA) efficiency 
			& \hspace{0.12cm}$0.39$ \\ 
			\hspace{0.15cm}$C_0$ & \hspace{0.2cm}Energy per complex operation 
						& \hspace{0.12cm}$10^{-9}$ Joule \\ 
\hline
\end{tabular} 
\end{center}
\end{table}%


\subsection{Total Energy Consumption}

We characterize $E(R, \, \vartheta,t)$ in two parts: (i) computation energy consumption, $E_{\text{comp}}(R, \, \vartheta,t)$, and (ii) communication energy consumption, $E_{\text{comm}}(R, \, \vartheta,t)$, i.e.,
\begin{align}
\label{eq:etot}
E(R, \, \vartheta,t) & = E_{\text{comp}}(R, \, \vartheta,t) + E_{\text{comm}}(R, \, \vartheta,t) \, .
\end{align}
\vspace{-0.5 cm}

\subsubsection{Computation Energy Consumption}
The computation energy consumption of a user's offloaded task depends on the task size, effective switching capacitance of the server, and its operating CPU frequency \cite{Quek, Kuper}. For the general $\I$ types of user scenario, we assume that the CS and MEC servers have different operating frequencies for each type of users. When $f_{\text{mec},i}$ and $f_{\text{cs},i}$ are, respectively, the operating CPU frequencies of MEC servers and CS for the $i^{\text{th}}$ type of user, the overall average computation energy consumption is given by
\begin{align}
\label{eq:ecomp}
\nonumber E_{\text{comp}} (R, \, \vartheta,t) & = \vartheta \kappa_c \sum\limits_{i=1}^{\I} p_i  f_{\text{cs},i}^\delta \bar{T}_{\text{c},i}\\
& + (1-\vartheta)\kappa_m \sum\limits_{i=1}^{\I} p_i f_{\text{mec},i}^\delta \bar{T}_{\text{mec},i} \, ,
\end{align}
\noindent where $\bar{T}_{\text{c},i}$ and $\bar{T}_{\text{mec},i}$ , respectively, represent the average computation time for the $i^{\text{th}}$ type of task at the CS and at the MEC servers, and $\delta$ is a constant (usually close to $3$) \cite{Kuper}. Note that $\bar{T}_{\text{c},i} = \frac{1}{\mu_{\text{c},i}}$ and $\bar{T}_{\text{mec},i} = \frac{1}{\mu_{\text{m},i}}$, where $\mu_{\text{m},i}$ and $\mu_{\text{c},i}$ are defined in terms of the operating frequencies in Table~\ref{table:varMpu}. From the definitions of $\mu_{\text{m},i}$ and $\mu_{\text{c},i}$ in Table~\ref{table:varMpu}, we observe that $E_{\text{comp}}(R, \, \vartheta,t)$ monotonically increases with $\vartheta$, if $\frac{f_{\text{cs},i}}{f_{\text{mec},i}}> \big({\kappa_m}/{\kappa_c}\big)^{1/(\delta-1)}$, for all $i = 1, 2, \ldots, \I$.

\begin{table}[!t]
\caption{Computation Energy Consumption Parameters \label{table:varMpu}} 
\begin{center}
\rowcolors{2}
{cyan!15!}{}
\renewcommand{\arraystretch}{1.5}
\begin{tabular}{l l || l l}
\hline 
 {\bf Parameters} & {\bf Values} & {\bf Parameters} & {\bf Values} \\
\hline 
\hspace{0.cm}$f_{\text{mec},1}$ & \hspace{0cm}$1$ GHz
			& \hspace{0.cm}$f_{\text{mec},2}$ & \hspace{0cm}$3.4$ GHz \\ 
\hspace{0.cm}$f_{\text{cs},1}$ & \hspace{0cm}$4$ GHz 
			& \hspace{0.cm}$f_{\text{cs},2}$ & \hspace{0cm}$5$ GHz  \\ 
\hspace{0.cm}$\kappa_m$  & \hspace{0cm}$10^{-27}$ J/cycle 
			& \hspace{0.cm}$\kappa_c$  & \hspace{0cm}$10^{-26}$ J/cycle  \\ 
\hspace{0.cm}$L_u$  & \hspace{0cm}$0.5$ Mbits 
			& \hspace{0.cm}$L_d$  & \hspace{0cm}$0.05$ Mbits \\ 
\hspace{0.cm}$C_p$ & \hspace{0cm}$330$ Cycles/byte
			& \hspace{0.cm}$\delta$  & \hspace{0cm}$3$ \\ 
\hspace{0.cm}$\mu_{m,1}$  & \hspace{0cm}$\frac{8f_{\text{mec},1}}{C_p L_u}$
			& \hspace{0.cm}$\mu_{m,2}$  & \hspace{0cm}$\frac{8f_{\text{mec},1}}{C_p L_u}$ \\ 
\hspace{0.cm}$\mu_{c,1}$ & \hspace{0cm}$\frac{8f_{\text{cs},1}}{C_p L_u}$ 
			& \hspace{0.cm}$\mu_{c,2}$ & \hspace{0cm}$\frac{8f_{\text{cs},1}}{C_p L_u}$ \\ 
\hline
\end{tabular} 
\vspace{- 0.5cm}
\end{center}
\end{table}%

\subsubsection{Communication Energy Consumption}

The communication energy consumption of a user in the network consists of the transmission energy consumption and the power consumption due to various circuit power consumption parameters (PCPs) \cite{Mohammed, HoydisEmil, Auer, Zander}. We have listed these various PCPs in Table~\ref{table:Eparam}. We also note that the average number of APs connected to a user is $\lambda_b \pi R^2$ (since each user connects to all APs within a coverage area, a circle of radius $R$). For the uplink communication scenario, the overall power consumption is therefore given by
\begin{align}
\label{eq:ulpower}
\nonumber P_{\text{ul}} &  =  \lambda_b \pi R^2 \left[ M(P_{\text{b}} + 2 \, C_0 \, B) + B \, r_u \, P_{\text{dec}} \right] + P_{\text{ue}} + P_{\text{d}} \\
& \hspace{4 cm} + \, B \, r_u \, P_{\text{cod}} + P_u \zeta \, .
\end{align}
\noindent Here $B$ is the communication bandwidth, $2M B$ is the total number of complex operations required in a single MRC detection \cite{Mohammed, HoydisEmil, Vandenberghe}, $P_{\text{ue}}$ is the fixed circuit power consumption at the transmitting end (i.e. user), and $r_u = \log_2(1+\widehat{\gamma})$ is the threshold spectral efficiency for uplink. In the similar fashion, the total downlink power consumption is given by
\begin{align}
\label{eq:dlpower}
\nonumber P_{\text{dl}} & =  \lambda_b \pi R^2 \left[ \zeta \, P_t + P_{\text{bs}} + P_{\text{osc}} + M(P_{\text{b}} + 2 \, C_0 \, B) \right.\\
& \hspace{2 cm} \left. + B \, r_d \, P_{\text{cod}} \right] + P_{\text{dec}} B r_d + P_{\text{d}} \, ,
\end{align}
\noindent where $2M B$ is the number of complex operations required for conjugate beamforming, $r_d = \log_2(1+\widetilde{\gamma})$ is the target spectral efficiency for downlink, and $P_t$ is the average downlink transmit power of AP. When the task size of a user in uplink is $L_u$ bits, the uplink transmission time is be given by $T_{\text{ul}} = \frac{L_u}{r_u B}$ seconds. Similarly, for the processed data size of $L_d$ bits in the downlink, the downlink transmission time is $T_{\text{dl}} = \frac{L_d}{r_d B}$ seconds. Therefore, the total communication energy consumption is given by
\begin{align}
\label{eq:ecom}
E_{\text{comm}}(R, \, \vartheta,t) & = P_{\text{ul}} \frac{L_u}{r_u B}+ P_{\text{dl}} \frac{L_d}{r_d B} \, .
\end{align}
\indent From \eqref{eq:ecomp}-\eqref{eq:ecom}, we observe that $E_{\text{comp}}(R, \, \vartheta,t)$  depends only on $\vartheta$, while $E_{\text{comm}}(R, \, \vartheta,t)$ monotonically increases on $R$. Therefore in order to solve the problem in \eqref{eq:mineconsumption}, we need to choose the minimum $R$ that satisfies the SECP constraint, i.e., $R = R^{\star}$, such that $p_{\text{secp}}(R^{\star}, \vartheta, t) = \xi$. Following this, we can now choose $\vartheta^{\star} = \vartheta(R^{\star})$, so that $E(R, \, \vartheta,t) \geq E(R^{\star}, \, \vartheta^{\star},t)$, for all $(R, \vartheta)$.

\subsection{Impact of SECP on $E(R^{\star}, \, \vartheta^{\star},t)$}

\begin{figure}[t!]
\vspace{-0.7 cm}
\hspace{0.1 cm}
    \begin{center}   
    { 
	\psfragscanon
	 \includegraphics[width=1.00\columnwidth]{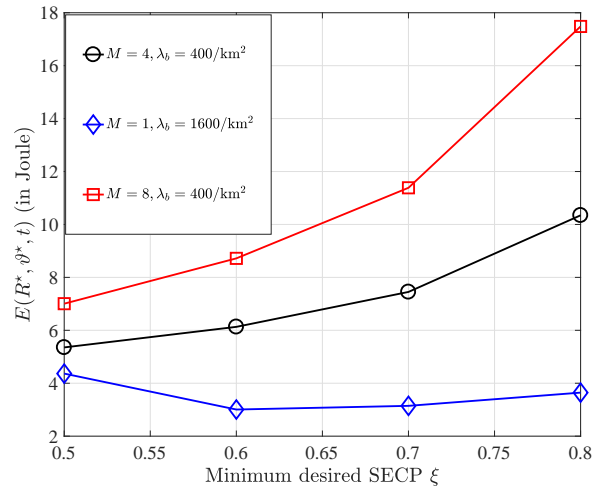}
    }
    \end{center}
    \caption{
    		Minimum required total average energy consumption, $E(R^{\star}, \, \vartheta^{\star},t)$, as a function of minimum desired SECP, $\xi$.	
		 }\vspace{-0.6 cm}
   \label{fig:evslatcdf}
\end{figure}

In Fig.~\ref{fig:evslatcdf}, we finally plot the minimum required total average energy consumption, $E(R^{\star}, \, \vartheta^{\star},t)$, as computed from solving the optimization problem in \eqref{eq:mineconsumption}, as a function of the minimum desired SECP, $\xi$, for the computation scenario in Fig.~\ref{fig:scpph1new}. For this figure, we use the following system parameter values: $\alpha = 3.7$, $B = 1$ MHz, $\lambda_d = 100$/km\textsuperscript{2}, target computation latency $t = 12$ ms, and $\widehat{\gamma} = \widetilde{\gamma} = 2.62$ dB (i.e. the target information rate for uplink and downlink transmission is $1.5$ bps/Hz). The values of the PCPs and computation parameters are provided in Table~\ref{table:Eparam} and Table~\ref{table:varMpu} respectively \cite{Auer, HoydisEmil, Mohammed, Quek, Nurminen}. From Table~\ref{table:varMpu}, we also note that $\frac{f_{\text{cs}}}{f_{\text{mec}}}> \big({\kappa_m}/{\kappa_c}\big)^{1/(\delta-1)}$, i.e., $E_{\text{comp}}(R, \, \vartheta,t)$ monotonically increases with $\vartheta$, for a given $R$ and $t$. Clearly, to compute $E(R^{\star}, \, \vartheta^{\star},t)$, the required $\vartheta^{\star}$ is given by
\begin{align}
\label{eq:reqtheta}
\vartheta^{\star} & = \min_{p_{\text{secp}}(R^{\star},\vartheta,t) \geq \xi} \vartheta(R^{\star}) \, .
\end{align}
\indent From Fig.~\ref{fig:evslatcdf}, we observe that the total energy consumption, $E(R^{\star}, \, \vartheta^{\star},t)$, first decreases and then increases with $\xi$ (see the blue curve with diamonds). This is due to the fact that with sufficiently large $\xi$, the increase in the required $R^{\star}$ with $\xi$ is significantly large and therefore $E_{\text{comm}}(R^{\star}, \, \vartheta^{\star},t)$ increases rapidly with $\xi$. We also observe that the required $\vartheta^{\star}$ decreases with $R^{\star}$ for the same value of the SECP in this regime (see Fig.~\ref{fig:secpph10}), and therefore, the computation energy consumption, $E_{\text{comp}}(R^{\star}, \, \vartheta^{\star},t)$, decreases with $\xi$ and converges to a small value. Consequently, in the large $\xi$ regime, $E_{\text{comm}}(R^{\star}, \, \vartheta^{\star},t)$ dominates the total energy consumption and therefore, $E(R^{\star}, \, \vartheta^{\star},t)$ increases with $\xi$.

On the other hand, when $\xi$ is sufficiently small, the required $R^{\star}$ is also small and consequently, the required $\vartheta^{\star}$ that satisfies the SECP constraint is significantly high (see Fig.~\ref{fig:secpph10}). Since, in this regime, the increase in $R^{\star}$ with the minimum desired SECP, $\xi$, is small, the overall increase in $E_{\text{comm}}(R^{\star}, \, \vartheta^{\star},t)$ is also small. On the other hand, due to large value of $\vartheta^{\star}$, the computation energy consumption, $E_{\text{comp}}(R^{\star}, \, \vartheta^{\star},t)$, is significantly large in this regime. Consequently, in the small $\xi$ regime, $E(R^{\star}, \, \vartheta^{\star},t)$ is dominated by $E_{\text{comp}}(R^{\star}, \, \vartheta^{\star},t)$ and since $\vartheta^{\star}$ rapidly decreases with $\xi$, $E_{\text{comp}}(R^{\star}, \, \vartheta^{\star},t)$ also decreases significantly. Consequently, $E(R^{\star}, \, \vartheta^{\star},t)$ is observed to decrease with $\xi$ in this regime. We also note that for a fixed AP density, $\lambda_b$, $E(R^{\star}, \, \vartheta^{\star},t)$ increases with $M$ (see the curve with red squares and the curve with black circles). This is due to the fact that the impact of $M$ being negligible on the SECP, the computation energy consumption does not change significantly with $M$, while the communication energy consumption increases linearly with $M$.

Next, we analyze the impact of AP density for a fixed antenna density, $M \lambda_b = 1600$/km\textsuperscript{2}, in the system (see the curve with blue diamonds and the curve with black circles). We observe that $E(R^{\star}, \, \vartheta^{\star},t)$ with higher AP density (i.e., $\lambda_b = 1600$/km\textsuperscript{2}) is smaller than the scenario with lower AP density (i.e., $\lambda_b = 400$/km\textsuperscript{2}) and this gap in $E(R^{\star}, \, \vartheta^{\star},t)$ increases with $\xi$. This shows that in edge computing-enabled CF massive MIMO systems, it is more energy efficient, if we deploy APs with higher density and small number of antennas, instead of deploying APs with smaller density and with larger number of antennas.


\section{Conclusion}\label{sec:conclusion}


{In this paper, we introduce novel edge computing-enabled cell-free massive MIMO systems. Considering the presence of computing servers at each AP and cloud server at the CS of the system, we firstly devise suitable communication and task computation strategies and analyze their respective performances in terms of the successful edge computing probability (SECP). Finally, using the SECP, we numerically characterize the total minimum energy consumption, required for guaranteeing a certain level of SECP. It is observed that for a given SECP, the energy consumption is minimized when we have higher AP density, i.e., it is more energy efficient to have additional APs, instead of having more antennas at each AP. The outcomes of our work provides insights into the energy efficient design of the MEC enabled CF massive MIMO systems, and also opens several issues for future work, including transmission power control for our joint transmission strategy in the downlink, performance analysis with other forms of cooperation strategies (e.g., collaborative beamforming) etc.}


\begin{appendix}

\subsection{Proof of Theorem~\ref{trm:outageul}}\label{app:outageproof}
From the expression of outage probability in \eqref{eq:outul}, we have
\begin{align}
\label{eq:outageul1}
\poutul(R) & \mya \sum\limits_{n=0}^{\infty}\frac{\bar{N}^n}{n!}e^{-\bar{N}}  \Prb \Big[ \max\limits_{k \in \{1,2,\ldots,n\} } \text{SIR}_{k0, \text{ul}} < \widehat{\gamma} | N = n \Big] \, ,
\end{align}
\noindent where step $(a)$ follows from the fact that the number of APs connected to the $0$-th user is a Poisson random variable, with mean $\bar{N} \Define \Eb{N} = \lambda_b \pi R^2$. Assuming that the SIR at different APs are independent, we have
\begin{align}
\label{eq:outageul2}
\nonumber \Prb \Big[ \max\limits_{k \in \{1,2,\ldots,n\} } \text{SIR}_{k0, \text{ul}} < \widehat{\gamma} | N = n \Big] & = \prod\limits_{k=1}^{n} \Prb[\text{SIR}_{k0, \text{ul}} < \widehat{\gamma}] \\
& = (1 - p_0(R))^n \, ,
\end{align}
\noindent where $p_0(R) \Define \Prb[\text{SIR}_{k0, \text{ul}} \geq \widehat{\gamma}] = 1 - \Prb[\text{SIR}_{k0, \text{ul}} < \widehat{\gamma}]$. Using \eqref{eq:outageul2} in \eqref{eq:outageul1} we have
\begin{align}
\label{eq:outageul3}
\poutul(R) & = \sum\limits_{n=0}^{\infty} \frac{(\lambda_b \pi R^2)^n}{n!} e^{-\lambda_b \pi R^2} (1 - p_0(R))^n \, = e^{-\lambda_b \pi R^2 p_0(R)} \, .
\end{align}
\par Using the definition of $\text{SIR}_{k0, \text{ul}}$ from \eqref{eq:sirul} in the definition of $p_0(R)$, we have
\begin{align}
\label{eq:covp}
\nonumber p_0(R) & = \int_{0}^{R}f_{x_{k0}}(r) \Prb[g_{k0} \geq l^{-1}(r)\widehat{\gamma} I | x_{k0} = r] \, dr\\
\nonumber & \myb \int_{0}^{R}f_{x_{k0}}(r) \Eb{\frac{\Gamma(M, l^{-1}(r)\widehat{\gamma}I)}{\Gamma(M)}} \, dr\\
 & \myc \sum\limits_{m=0}^{M-1}\frac{(-1)^m}{m!}\int_{0}^{R}f_{x_{k0}}(r) l^{-m}(r) \mathcal{L}^{(m)}(\widehat{\gamma}l^{-1}(r)) \, dr \, ,
\end{align}
\noindent where $f_{x_{k0}}(r)$ denotes the distribution of the link distance between the $0$-th user and the $k^{\text{th}}$ AP and $I \Define \sum_{q \in \bm \Phi_d\backslash \{0\} }\widetilde{g}_{kq} l(x_{kq})$ is the total interference power. Note that the user distribution follows a homogeneous PPP and the $k^{\text{th}}$ AP can connect to all users within a radius $R$. Therefore, $f_{x_{k0}}(r)$ is given by 
\begin{align}
\label{eq:distpdf}
f_{x_{k0}}(r) & =  \left \{ \begin{array}{ll}
\frac{2r}{R^2}, & 0 \leq r \leq R,\\
0, & \text{elsewhere}.
\end{array} \right.
\end{align}
\indent Furthermore, in \eqref{eq:covp}, the step $(b)$ follows from the fact that $g_{k0}$ is $\Gamma(M,1)$ distributed and the step $(c)$ follows from the following equality
\begin{align}
\label{eq:gammarel}
\Gamma(w,x) & = \Gamma(w) \sum\limits_{k=0}^{w-1}\frac{x^k}{k!}e^{-x} \, ,
\end{align}
\noindent where $w \in \Z^{+}$. Note that in the last line of \eqref{eq:covp}, we have $\mathcal{L}(s) \Define \Eb{e^{-s I}}$, the Laplace transform of the total interference power, $I$. Using \eqref{eq:covp} and \eqref{eq:distpdf} in \eqref{eq:outageul3} we obtain \eqref{eq:pout}. 

\par Finally, to compute $\mathcal{L}(s)$, using the PGFL \cite{Mecke}, we have
\begin{align}
\label{eq:lapdef}
\nonumber \mathcal{L}(s) & = \Eb{e^{-s I}} \, = \, \Eb{\prod\limits_{q \in \bm \Phi_d\backslash \{0\}} e^{-s \widetilde{g}_{kq}}}\\
\nonumber & = \exp \left \{ -2\pi \lambda_d \int_{0}^{\infty} \Big(1 - \frac{1}{1 + s l(r)} \Big) r \, dr \right\}\\
\nonumber & = \exp \left \{ -\pi \lambda_d \int_{0}^{\infty} \Big(1 - \frac{1}{1 + s l(\sqrt{z})} \Big) \, dz \right\}\\
\nonumber & = F_1(s) \exp \left \{ -\pi \lambda_d \int_{d_0^2}^{\infty} \Big(1 - \frac{1}{1 + s z^{-\frac{\alpha}{2}}} \Big) \, dz \right\}\\
& = F_1(s) \exp \left \{ -\pi \lambda_d \int_{d_0^2}^{\infty}  \frac{1}{1 + \frac{1}{s} z^{\frac{\alpha}{2}}}  \, dz \right\} \, = F_1(s) \, F_2(s) \, ,
\end{align}
\noindent where $F_1(s)$ is given by
\begin{align}
\label{eq:f1def}
\nonumber F_1(s) & = \exp \left \{ -\pi \lambda_d \int_{0}^{d_0^2} \Big(1 - \frac{1}{1 + s l(\sqrt{z}) } \Big) \, dz \right\}\\
\nonumber & = \exp \left \{ -\pi \lambda_d \int_{0}^{d_0^2} \Big(1 - \frac{1}{1 + s d_0^{-\alpha} } \Big) \, dz \right\}\\
& = e^{-\pi \lambda_d d_0^2 \Big (1 - \frac{1}{1+s d_0^{-\alpha} } \Big ) } \,. 
\end{align}
\indent Similarly, $F_2(s)$ is given by
\begin{align}
\label{eq:f2def}
\nonumber F_2(s) & = \exp \left \{ -\pi \lambda_d \int_{d_0^2}^{\infty}  \frac{1}{1 + \frac{1}{s} z^{\frac{\alpha}{2}}}  \, dz \right\}\\
& \myd \exp \left \{ -\frac{ \frac{2\pi \lambda_d}{\alpha} s d_0^{2 - \alpha}}{1 - \frac{2}{\alpha}} \HGF{2}{1}{1,1-\frac{2}{\alpha};2-\frac{2}{\alpha};-s d_0^{-\alpha}} \right\}
\end{align}
\noindent where step $(d)$ follows from the result $3.194$ in \cite{Ryzhik}. Differentiating \eqref{eq:lapdef} on both sides $m$ times, with respect to $s$, we obtain \eqref{eq:ulltderiv}.

\par Using \eqref{eq:f1def}, we can easily compute the $m^{\text{th}}$ order derivatives of $F_1(s)$ as given in \eqref{eq:partderiv1}. Similarly, using the definition of $F_2(s)$ in \eqref{eq:f2def} and differentiating it with respect to $s$, we have
\begin{align}
\label{eq:partderiv21}
\nonumber F^{(m)}_2(s) & = \pi \lambda_d \sum\limits_{i=0}^{m-1} {m-1 \choose i}(-1)^{m-i} (m-i)! F^{(i)}_2(s)\\
\nonumber & \hspace{2 cm} \times \underbrace{\int_{d_0^2}^{\infty}\frac{(z^{\frac{-\alpha}{2}})^{m-i}}{(1+s z^{\frac{-\alpha}{2}})^{m-i+1}} \, dz}_{\Define \, k_{m-i}}\\
& = \pi \lambda_d \sum\limits_{i=1}^{m} {m-1 \choose m-i}(-1)^{i} i! F^{(m-i)}_2(s) k_i \, ,
\end{align}
\noindent where $k_{m} = \int_{d_0^2}^{\infty}\frac{(z^{\frac{-\alpha}{2}})^{m}}{(1+s z^{\frac{-\alpha}{2}})^{m+1}} \, dz = \frac{2}{\alpha} d_0^{2 - m\alpha} \Big( m-\frac{2}{\alpha} \Big )^{-1} \HGF{2}{1}{m+1,m-\frac{2}{\alpha}; m-\frac{2}{\alpha}+1; - s d_0^{-\alpha}}$. Note that $\mathcal{L}^{(m)}(s)$ is given by \eqref{eq:ulltderiv}, which can be computed using \eqref{eq:partderiv1} and \eqref{eq:f1def}-\eqref{eq:partderiv21}.
%
%


\subsection{Proof of Theorem~\ref{trm:outagedl}}\label{app:outageproof2}
From \eqref{eq:outdl}, we have
\begin{align}
\label{eq:outagedl}
\nonumber \poutdl(R) & = \Prb\left[\frac{P}{I_{\text{dl}}} \leq \widetilde{\gamma}\right] = \Prb\left[I_{\text{dl}} \geq \frac{P}{\widetilde{\gamma}}\right] \\
\nonumber & \mya \Eb{\frac{\Gamma(\zeta(R), \frac{P}{\eta \widetilde{\gamma}})}{\Gamma(\zeta(R))}}\\
& \, \mbox{\scriptsize $ {\gleq \atop {k_0 = \lceil \zeta(R) \rceil}}$ } \sum\limits_{m=0}^{k_0-1}\frac{(-1)^m (\widetilde{\gamma} \eta)^{-m}}{m!}\mathcal{L}_P^{(m)}\Big( \frac{1}{\widetilde{\gamma} \eta}\Big) \, ,
\end{align}
\noindent where step $(a)$ follows from the approximation of $I_{\text{dl}}$ in Proposition~\ref{intfcharac}, and the last step follows from \eqref{eq:gammarel}. Here $\mathcal{L}_P^{(m)}(s)$ is defined in Theorem~\ref{trm:outagedl}.

\par Now, from the definition of $\mathcal{L}_P(s)$ we have $\mathcal{L}_P(s) = e^{-2\pi \lambda_b \varrho(s)}$, where $\varrho(s)$ is given by
\begin{align}
\label{eq:sigLap}
\nonumber \varrho(s) & \Define \int\limits_{0}^{R} \Big( 1- \Eb{e^{-s \, g \, l(r) }} \Big) r \, dr \\
\nonumber & = \int\limits_{0}^{\infty}\int\limits_{0}^{R} (1 - e^{-s \, g \, l(r) } ) r \, dr \, f_g(g) \, dg\\
\nonumber & = \frac{1}{2}R^2 - \frac{d_0^2}{2}\Eb{e^{-s \, g \, d_0^{-\alpha} } } - \Eb{\int_{d_0}^{R}r \, e^{-s \, g \, r^{-\alpha} } \, dr}\\
\nonumber & = \frac{1}{2}R^2 - \frac{d_0^2}{2}\Eb{e^{-s \, g \, d_0^{-\alpha} } } \\
\nonumber & \hspace{1 cm} - \Eb{\frac{1}{\alpha} (s \, g)^{\frac{2}{\alpha}} \int_{s \, g \, R^{-\alpha}}^{s \, g \, d_0^{-\alpha}}e^{-u}u^{-1-\frac{2}{\alpha}}\, du}\\
\nonumber & = \frac{1}{2}R^2 - \frac{d_0^2}{2}\Eb{e^{-s \, g \, d_0^{-\alpha} } }\\
\nonumber & \hspace{1 cm} -  \frac{1}{2}\Eb{R^2 \, e^{-s \, g \, R^{-\alpha}} - d_0^2e^{-s \, g \, d_0^{-\alpha}}}\\
\nonumber & \hspace{1 cm}+  \frac{1}{2} \Eb{(s \, g)^{\frac{2}{\alpha}}\int_{s \, g \, R^{-\alpha}}^{s \, g \, d_0^{-\alpha}}e^{-u}u^{-\frac{2}{\alpha}}\, du \Big)}\\
\nonumber & = \frac{1}{2}\Eb{R^2(1 - e^{-s g R^{-\alpha}})}\\
\nonumber &  + \frac{1}{2}\Eb{(s \, g)^{\frac{2}{\alpha}} \Big\{ \Gamma(1-\frac{2}{\alpha}, s g R^{-\alpha}) - \Gamma(1-\frac{2}{\alpha}, s g d_0^{-\alpha})\Big\}}\\
\nonumber & = \frac{1}{2}{R^2 \Bigg(1 - \frac{1}{(1+sR^{-\alpha})^M} \Bigg)}\\
\nonumber & \hspace{1 cm}   + \frac{1}{2}s^{\frac{2}{\alpha}} \E \left[ g^{\frac{2}{\alpha}} \left\{ \Gamma \left(1-\frac{2}{\alpha}, s g R^{-\alpha}\right) \right. \right. \\
& \hspace{3.3 cm} \left. \left. - \, \Gamma \left(1-\frac{2}{\alpha}, s g d_0^{-\alpha}\right)\right\} \right] \, .
\end{align}
\indent Note that $\mathcal{L}_P(s)$ and its $m^{\text{th}}$ order derivatives are given by \eqref{eq:lapdevm}, where $\varrho^{(m)}(s)$ denotes the $m^{\text{th}}$ order derivative of $\varrho(s)$. Using the definition of $\varrho(s)$ in \eqref{eq:sigLap}, we have
\begin{align}
\label{eq:gamedevm}
\nonumber \varrho^{(m)}(s) & = (-1)^{m-1}   \Eb{g^m \int\limits_{0}^{R} r l^m(r) e^{-s \, g \, l(r)} \, dr}\\
\nonumber & = (-1)^{m-1} \Eb{g^m  \frac{1}{2}d_0^{2-m\, \alpha}e^{-s \, g \, d_0^{-\alpha}}}\\
\nonumber & + (-1)^{m-1} \Eb{g^m \int\limits_{d_0}^{R} r^{-m \, \alpha + 1} \, e^{-s \, g r^{-\alpha}} \, dr}\\
\nonumber & = (-1)^{m-1}\frac{1}{2}d_0^{2-m \alpha}\frac{\Gamma(M+m)}{\Gamma(M) (1+sd_0^{-\alpha})^{M+m} }\\
& + \frac{1}{\alpha} \Eb{g^m H(R, d_0, \alpha, s)}
\end{align}
\noindent where $H(R, d_0, \alpha, s) = (s \, g)^{\frac{2}{\alpha}-m} \Big[ \Gamma(m-\frac{2}{\alpha}, s g R^{-\alpha}) - \Gamma(m-\frac{2}{\alpha}, s g d_0^{-\alpha}) \Big]$.

\subsection{Proof of Theorem~\ref{trm:scp}}\label{app:scpmg1}

Using the definition of $T_{\text{comp}}$ from \eqref{eq:complat}, in \eqref{eq:scpdef}, we have
{\begin{align}
\label{eq:comp1}
\nonumber \Prb[T_{\text{comp}} \leq t] & = \vartheta \Prb[T_{\text{comp}} \leq t | \bm E_1]\\
\nonumber & + (1-\vartheta)\Prb[T_{\text{comp}} \leq t | \bm E_1^c]\\
\nonumber & = \vartheta \Prb[\Tcs \leq t] + (1-\vartheta)\Prb[T_{\text{mec}} \leq t]\\
& = \vartheta \int\limits_{0}^{t} f_{\Tcs}(u) du + (1-\vartheta)\Prb[T_{\text{mec}} \leq t] \,,
\end{align}}
\noindent {where $\bm E_1$ is the event that the computation takes place at the CS, $\Prb[\bm E_1] = \vartheta$ and $\Prb[\bm E_1^c] = 1- \vartheta$. Here, the pdf of $\Tcs$, $f_{\Tcs}(u)$,  is computed from its Laplace transform, obtain from the P-K formula of M/G/1 queues \cite{Kleinrock}, as} 
{\begin{align}
\label{eq:Lapcs}
f_{\Tcs}(u) & = \mathcal{L}_{\Tcs}^{-1}\left[ (1- \rho_{\text{c}})\frac{s\mathcal{L}_{\tau_{\text{c}}}(s)}{s-\Lc + \Lc \mathcal{L}_{\tau_{\text{c}}}(s) } \right] \, ,
\end{align}}
\hspace{-0.3 cm} \noindent {where $\mathcal{L}_{\tau_{\text{c}}}(s) = \sum\limits_{i=1}^{\mathcal{I}} \frac{p_i \mu_{\text{c},i}}{s+\mu_{\text{c},i}}$ is the Laplace transform of the service time distribution at the CS (see \eqref{eq:servtpdf}). Using this in \eqref{eq:Lapcs}, we obtain \eqref{eq:tcsi}.}
\par {In \eqref{eq:comp1}, $\Prb[\Tmec \leq t]$ can be given as}
{\begin{align}
\label{eq:mec01}
\nonumber \Prb[T_{\text{mec}} \leq t] & = \sum\limits_{n=1}^{\infty} \Prb[N = n]\Prb[T_{\text{mec}} \leq t | N = n]\\
 & \mya \sum\limits_{n=1}^{\infty}\frac{(\lambda_b \pi R^2)^n}{n!}e^{-\lambda_b \pi R^2}\Prb[T_{\text{mec}} \leq t | N = n]  \, ,
\end{align}}
\hspace{-0.3 cm} {\noindent where $(a)$ is from the Poisson distribution of $N$, with mean $\lambda_b \pi R^2$. Since in case of processing at the MEC servers, the task is processed at the server with minimum load (i.e., minimum queue length), we define the minimum queue length among $n$ connected MEC servers, $\bar{N}_{\text{m}}(n)$, as follows}
{\begin{align}
\label{eq:minQlen}
\bar{N}_{\text{m}}(n) & = N_{\text{m},\widehat{k}}\,, \text{where} \,\, \widehat{k} \Define \arg \min\limits_{k \in \{1,2,\ldots,n\} } N_{\text{m},k} \, ,
\end{align}}
\hspace{-0.3 cm} {\noindent where $N_{\text{m},k}$ is the queue length of the $k^{\text{th}}$ MEC server. From \eqref{eq:minQlen}, it is clear that $\Tmec = T_{\text{m},\widehat{k}} = \sum\limits_{l=1}^{\bar{N}_{\text{m}}(n)+1} \tau_{\text{m},\widehat{k},l}$, where $\tau_{\text{m},\widehat{k},l}$ is the service time for the $l^{\text{th}}$ task at the $\widehat{k}$-th MEC server, for a given $\bar{N}_{\text{m}}(n)$. Clearly, $\Prb[T_{\text{mec}} \leq t | N = n]$ is given by}

%
%
{\begin{align}
\label{eq:mec1}
\nonumber \Prb[T_{\text{mec}} \leq t | N = n] & = \sum\limits_{v=0}^{\infty}\Prb[\bar{N}_{\text{m}}(n) = v]\\
& \hspace{1 cm} \times \,  \int\limits_{0}^{t} f_{\Tmec | \bar{N}_{\text{m}}(n) = v }(u) \, du   \, ,
\end{align}}
\hspace{-0.15 cm}{\noindent where $f_{\Tmec | \bar{N}_{\text{m}}(n) = v }(u)$ denotes the probability density function of $\Tmec$ for a given $\bar{N}_{\text{m}}(n)$ and it is computed as} 
{\begin{align}
\label{eq:tmecLT}
\nonumber f_{\Tmec | \bar{N}_{\text{m}}(n) = v }(u) & = \mathcal{L}_{\Tmec | \bar{N}_{\text{m}}(n) = v}^{-1}\left[ \prod\limits_{l=1}^{v+1} \mathcal{L}_{\tau_{\text{m},\widehat{k},l}}(s) \right]\\
& \myb  \mathcal{L}_{\Tmec | \bar{N}_{\text{m}}(n) = v}^{-1}\left[ \left(\sum\limits_{i=1}^{\I} \frac{p_i \mu_{\text{m},i} }{s + \mu_{\text{m},i} } \right)^{v+1} \right] \, .
\end{align}}
\hspace{-0.3 cm}{\noindent where (b) follows from \eqref{eq:servtpdf}. To compute $\Prb[\bar{N}_{\text{m}}(n) = v]$, we define it as}
{\begin{align}
\label{eq:tqpmf}
\Prb[\bar{N}_{\text{m}}(n) = v)] & = \Prb[\bar{N}_{\text{m}}(n) \leq v)] - \Prb[\bar{N}_{\text{m}}(n) \leq v-1)] \, .
\end{align}}
\hspace{-0.3 cm}{\noindent To evaluate \eqref{eq:tqpmf}, we first compute the moment generating function (MGF) of $N_{\text{m},k}$ using the P-K formula \cite{Kleinrock}, as}
{\begin{align}
\label{eq:mec13}
G_{\text{m},k}(z) & = \frac{(1-\rho_{\text{m}})(1-z)\mathcal{L}_{\tau_{\text{m}}}(\Lm - \Lm z)}{\mathcal{L}_{\tau_{\text{m}}}(\Lm - \Lm z) - z}= \sum\limits_{i=1}^{\I}\frac{\epsilon_i}{1-\omega_i z}\, ,
\end{align}}
\hspace{-0.3 cm}{where $\mathcal{L}_{\tau_{\text{m}}}(s)$ is computed using \eqref{eq:servtpdf} and $\epsilon_i$ and $\omega_i$, $\forall i = 1,2,\ldots, \I$, are defined in \eqref{eq:mec131} and \eqref{eq:mec132}, respectively. From \eqref{eq:mec13}, we have $\Prb[N_{\text{m},k} = v] = \sum\limits_{i=1}^{\I} \epsilon_i \omega_i^v$. From the definition of $\bar{N}_{\text{m}}(n)$, we have}
{\begin{align}
\label{eq:tqcdf}
\nonumber \Prb[\bar{N}_{\text{m}}(n) \leq v)] & = 1 - \Prb[\bar{N}_{\text{m}}(n) > v)] = 1 - \prod\limits_{k=1}^{n}\Prb[N_{\text{m},k} > v]\\
& = 1 - \left( \sum\limits_{i=1}^{\I} \frac{\epsilon_i \omega_i^{v+1}}{1 - \omega_i} \right)^n \, .
\end{align}}
\hspace{-0.3 cm} {\noindent Using \eqref{eq:tqcdf} in \eqref{eq:tqpmf}, we obtain $\Prb[\bar{N}_{\text{m}}(n) = v] = \left[\left( \sum\limits_{i=1}^{\I} \frac{\epsilon_i \omega_i^{v}}{1 - \omega_i} \right)^n - \left( \sum\limits_{i=1}^{\I} \frac{\epsilon_i \omega_i^{v+1}}{1 - \omega_i} \right)^n \right]$, and finally, using it in \eqref{eq:mec1}, we obtain \eqref{eq:mecPnew}.}

\end{appendix}

\ifCLASSOPTIONcaptionsoff
  \newpage
\fi



%

\bibliographystyle{IEEEtran}
\bibliography{IEEEabrvn,StringDefinitions,mybibn}

\end{document}